\documentclass[11pt]{article} 
\usepackage{fullpage}
\usepackage{graphicx}
\usepackage{upref}
\usepackage{enumerate}
\usepackage{latexsym}
\usepackage{pdfpages}
\usepackage[bookmarks]{hyperref}

\usepackage{color,graphics}
\usepackage{comment} 
\usepackage{caption}
\usepackage{subcaption}
\usepackage{wrapfig}
\usepackage{braket}
\usepackage{amssymb}
\usepackage{amsmath}
\usepackage{amsthm}
\usepackage{mathrsfs}
\usepackage{mathtools}
\usepackage{commath}
\usepackage{thmtools,thm-restate}
\usepackage{breqn}
\usepackage{boxedminipage}

\DeclareMathOperator{\cL}{\mathcal{L}}

\DeclareMathOperator{\basis}{\mathbf{B}}

\newcommand{\cc}[1]{\ensuremath{\mathsf{#1}}}

\newcommand{\NP}{\cc{NP}}

\newcommand{\coNP}{\cc{coNP}}

\newcommand{\NTIME}{\cc{NTIME}}

\newcommand{\NPPoly}{\cc{NP}/\cc{Poly}}

\usepackage{amsfonts}
\usepackage{varioref}
\usepackage[ansinew]{inputenc}
\usepackage[many]{tcolorbox}
\usepackage{thm-restate}

\usepackage{xcolor}
\hypersetup{
	colorlinks,
	linkcolor={red!75!black},
	citecolor={blue!75!black},
	urlcolor={blue!75!black}
}
\usepackage[hyperpageref]{backref}
\RequirePackage[capitalize,nameinlink,noabbrev]{cleveref}

\newtheorem{theorem}{Theorem}[section]
\newtheorem{lemma}[theorem]{Lemma}
\newtheorem{corollary}[theorem]{Corollary}

\newtheorem{defn}[theorem]{Definition}

\theoremstyle{remark}
\newtheorem{remark}{Remark}

\theoremstyle{definition}

\AtBeginDocument{}

\DeclareMathOperator{\real}{\mathbb{R}}
\DeclareMathOperator{\nat}{\mathbb{N}}

\newcommand{\intg}{\mathbb{Z}}
\newcommand{\ratn}{\mathbb{Q}}

\newcommand{\poly}{\mathrm{poly}}

\newcommand{\Poly}{\mathrm{Poly}}

\newcommand{\SVP}{\mathsf{SVP}}

\newcommand{\CVP}{\mathsf{CVP}}

\newcommand{\GapCVP}{\mathsf{GapCVP}}
\newcommand{\CVPip}{\mathsf{CVP}^{\mathsf{IP}}}

\newcommand{\CVPmvp}{\mathsf{CVP}^{\mathsf{mvp}}}

\newcommand{\GapCVPmvp}{\mathsf{GapCVP}^{\mathsf{mvp}}}

\newcommand{\sat}{\mathsf{SAT}}
\newcommand{\SAT}{\mathsf{SAT}}
\newcommand{\seth}{\mathsf{SETH}}
\newcommand{\SETH}{\mathsf{SETH}}

\newcommand{\cS}{{\mathcal S}}

\newcommand{\IP}{\mathsf{IP}}
\newcommand{\mvp}{\mathsf{mvp}}

\DeclareMathOperator{\proj}{proj}
\newcommand{\vect}[1]{\boldsymbol{#1}}
\newcommand{\eps}{\varepsilon}
\renewcommand{\epsilon}{\varepsilon}
\renewcommand{\vec}[1]{\vect{#1}}

\newcommand{\dist}{\mathrm{dist}}

\begin{document}

\title{Why we couldn't prove SETH hardness of the Closest Vector Problem for even norms!}
\author{}
\author{Divesh Aggarwal\\ CQT, National University of Singapore \\
\texttt{dcsdiva@nus.edu.sg} 
\and Rajendra Kumar
\\
Weizmann Institute of Science, Israel\\
\texttt{rjndr2503@gmail.com}
}
\date{}
\maketitle

\begin{abstract}
Recent work~\cite{BGS17,ABGS19} has shown $\SETH$ hardness of $\CVP$ in the $\ell_p$ norm for any $p$ that is not an even integer. This result was shown by giving a Karp reduction from $k$-$\SAT$ on $n$ variables to $\CVP$ on a lattice of rank $n$. In this work, we show a barrier towards proving a similar result for $\CVP$ in the $\ell_p$ norm where $p$ is an even integer. 
  We show that for any $c>0$, if for every $k > 0$, there exists an efficient reduction that maps a $k$-$\SAT$ instance on $n$ variables to a  $\CVP$ instance for a lattice of rank at most $n^{c}$ in the Euclidean norm, then $\coNP \subset \NPPoly$. We prove a similar result for $\CVP$ for all even norms under a mild additional promise that the ratio of the distance of the target from the lattice and the shortest non-zero vector in the lattice is bounded by $\exp(n^{O(1)})$.
  
  Furthermore, we show that for any $c > 0$, and any even integer $p$, if for every $k > 0$, there exists an efficient reduction that maps a $k$-$\SAT$ instance on $n$ variables to a  $\SVP_p$ instance for a lattice of rank at most $n^{c}$, then $\coNP \subset \NPPoly$.\footnote{The result for $\SVP$ does not require any additional promise.}

  While prior results have indicated that lattice problems in the $\ell_2$ norm (Euclidean norm) are easier than lattice problems in other norms, this is the first result that shows a separation between these problems. 

We achieve this by using a result by Dell and van Melkebeek~\cite{dell2014satisfiability} on the impossibility of the existence of a reduction that compresses an arbitrary $k$-$\SAT$ instance into a string of length $\mathcal{O}(n^{k-\epsilon})$ for any $\epsilon>0$. 

In addition to $\CVP$, we also show that the same result holds for the \textsf{Subset-Sum} problem using similar techniques. 
\end{abstract}

\thispagestyle{empty}
\newpage
\tableofcontents
\pagenumbering{roman}
\newpage
\pagenumbering{arabic}

\section{Introduction}

\subsection{Lattice Problems}

A lattice $\cL$ is the set of integer linear combination of $n$ linearly independent vectors $\vect{b}_1,\vect{b_2},\cdots,\vect{b}_n \in \real^m$, \textit{i.e.}
\[\cL(\vect{b}_1,\vect{b}_2,\cdots,\vect{b}_n):=\left\{\sum\limits_{i=1}^n z_i\vect{b}_i : \forall i\in [n], z_i\in \intg \right\}.\]
We call $n$ as the rank of the lattice, $m$ as the dimension of the lattice and $\basis=\{\vect{b}_1,\vect{b_2},\cdots,\vect{b}_n\}$ is a basis of the lattice. There exist many different basis of a lattice. 

The two most important computational problems on lattices are the Shortest Vector Problem ($\SVP$) and the Closest Vector Problem ($\CVP$). In the Shortest Vector Problem, given a basis for a lattice the goal is to output a shortest non-zero lattice vector. For $\gamma>1$, in the $\gamma$-approximation of $\SVP$ ($\gamma$-$\SVP$) the goal is to output a nonzero lattice vector whose length is at most $\gamma$ times the length of shortest non-zero lattice vector. In the Closest Vector Problem, given a target vector and basis for a lattice, the goal is to output a closest lattice vector to the target vector. For $\gamma>1$, in the $\gamma$-approximation of $\CVP$, ($\gamma$-$\CVP$) the goal is to output a lattice vector whose distance from target vector is at most $\gamma$ times the minimum distance between the target vector and lattice. In this work, we only consider the length and distance in the $\ell_p$ norms, defined as follows. For $1\leq p<\infty$
\[\|\vect{x}\|_p:=\left(\sum\limits_{i=1}^m |x_i|^p\right)^{1/p}\]
and for $p = \infty$,
\[\|\vect{x}\|_\infty :=\max\limits_{i=1}^m\{|x_i|\}.\]
We write $\SVP_p$ and $\CVP_p$ for the respective problems in $\ell_p$ norm. The most commonly used norm is the $\ell_2$ norm which is also called the Euclidean norm. The $\CVP$ is known to be at least as hard as $\SVP$ (in the same norm). More specifically, there is an efficient polynomial-time reduction from $\SVP_p$ to $\CVP_p$, which preserves the dimension, rank, and approximation factor~\cite{GMSS99}.

Computational problems on lattices are particularly important due to their connection to lattice-based cryptography. Most specifically, the security of many cryptosystems~\cite{Ajtai96,MR04,Regev09,Regev06,MR08,Gentry09,BV14} is based on the hardness of polynomial approximation of lattice problems. Other than the design of cryptosystems, from the 80's the solvers for lattice problems has its application in algorithmic number theory~\cite{LLL82}, convex optimization~\cite{Kannan87,FrankT87}, and cryptanalytic tools~\cite{Shamir84,Brickell84,LagariasO85}.

Algorithms for $\CVP$ have been extensively studied for a long time. Kannan gave an enumeration algorithm~\cite{Kannan87} for $\CVP$ that works for any norm and takes $n^{O(n)}$ time and requires $\poly(n)$ space. Micciancio and Voulgaris gave a deterministic algorithm for $\CVP_2$ that takes $2^{2n+o(n)}$ time and requires $2^{n+o(n)}$ space. Aggarwal, Dadush, and Stephens-Davidowitz~\cite{ADS15} gave the current fastest known algorithm for $\CVP_2$; it takes $2^{n+o(n)}$ time and space. However, there is no progress in solving exact $\CVP_p$ in $\ell_p$ norm; still, Kannan's algorithm is the fastest for exact $\CVP_p$ for any arbitrary $p$. For a small approximation of $\CVP_p$, Blomer and Seifert~\cite{BN09} gave an algorithm that runs in $2^{O(m)}$ time. Later, Dadush~\cite{dadush12} improved it by giving a $2^{O(n)}$ time algorithm. For $p=\infty$, Aggarwal and Mukhopadhyay~\cite{AM18} gave a $2^{2m+o(m)}$ time algorithm. Recently, Eisenbrand and Venzin~\cite{EV22}, gave an algorithm for a (large enough) constant-factor approximation of $\CVP$ for any $\ell_p$ norm in $2^{0.802m +o(m)}$ time.

The $\CVP$ in any norm is NP-hard~\cite{vEB81}. It is also known to be NP-hard for almost polynomial approximation $n^{c/\log \log n}$ for some constant $c>0$\cite{DKRSApproximatingCVP03,DinApproximatingSVPinfty02}. %
All of these hardness results do not say anything about the fine-grained hardness of lattice problems. In particular, it is not possible to say anything about possibility/impossibility of a $2^{\sqrt{n}}$ or a $2^{n/100}$ time algorithm for $\CVP$. All known cryptanalytic attacks on lattice-based cryptosystems proceed via solving near exact lattice problems in a small dimensional lattice, and so,  if one were to find an algorithm for $\CVP$ or $\SVP$ that runs in, say, $2^{{n}/100}$-time, then it will break all lattice-based cryptosystems currently considered to be practical. This immediately leads to the question \emph{whether such attacks can be ruled out by giving appropriate lower bounds for lattice problems under reasonable assumptions}. 

Motivated by the above question, Bennett, Golovnev, and Stephens-Davidowitz~\cite{BGS17} initiated the study of the fine-grained hardness of $\CVP$ and its variants. They showed that, for any constant $\epsilon>0$ and $p\not\in 2\intg$, if there exists an algorithm for $\CVP_p$ that runs in time $2^{(1-\epsilon)n}$ then it refutes the Strong Exponential Time Hypothesis ($\seth$). The authors additionally also showed that for any $p \ge 1$, there exists no algorithm for $\CVP_p$ that runs in time $2^{o(n)}$ under the Exponential Time Hypothesis. Later, these results were extended to other lattice problems~\cite{AS18b,AC21,BP20,BPT22}. Aggarwal, Bennett, Golovnev, and Stephens-Davidowitz~\cite{ABGS19}, improved the fine-grained hardness of $\CVP$ and showed that even approximating $\CVP_p$ to a factor slightly bigger than $1$ is not possible in time $2^{(1-\epsilon)n}$ under the Gap variant of the Strong Exponential Time Hypothesis. This result was again only shown for $p\not\in 2\intg$.  This immediately leads to the question whether such a hardness result is possible for $p \in 2 \intg$, or if there is a fundamental barrier that does not allow such a result. 
This is particularly important/interesting for the Euclidean norm (i.e., $p = 2$) since the security of lattice-based cryptosystems is typically based on the hardness of lattice problems in the Euclidean norm. The authors~\cite{ABGS19} made a small progress towards answering this question by proving that there are no ``natural" reductions from $k$-$\SAT$ on $n$ variables to $\CVP_2$ on a lattice of rank at most $4n/3$.\footnote{A reduction is said to be natural if there is a bijective mapping between the set of satisfying assignments, and the set of closest vectors in the lattice.}

Motivated by the fact that the computational problems in lattices run in time exponential in the dimension of the lattice,~\cite{ACKLS21} initiated the study of exponential time reductions (reductions that run in time $2^{\eps m}$, for some small constant $\eps$) between $\CVP$ and $\SVP$ in $\ell_p$ norms for different $p$. The techniques used to obtain the results in this work were based on~\cite{EV20}. Together, these results have shown that for large constant approximation factors, both $\CVP_p$ and $\SVP_p$ for any $p \ge 1$ are almost equivalent.  

Unfortunately, our main result (almost) rules out the possibility of an efficient reduction from $k$-$\SAT$ on $n$ variables to $\CVP_p$ for any even integer $p$ with rank bounded by a fixed polynomial in $n$. We achieve this result by proving that a $\CVP_p$ instance for any even integer $p$ can be compressed to size that is roughly a fixed polynomial in the rank of the lattice, while $k$-$\SAT$ is only known to be compressible in $O(n^k)$ bits.

\subsection{\textsf{Subset Sum} Problem}

Motivated by the difficulty in proving $\SETH$ hardness for $\CVP$ for even norms, we consider the closely related but relatively simpler \textsf{Subset Sum} problem. The \textsf{Subset Sum} problem is among the most fundamental computational problems described as follows. Given $n$ positive integers $x_1, \ldots, x_n$, and a target value $t$, the goal is to decide whether there exists a subset of the $n$ positive integers that sums to $t$. The classical algorithms for this problem run in time $O(tn)$ via dynamic programming~\cite{bellman1966dynamic}, and in time $2^{n/2} \cdot \poly(\log t, n)$ using meet-in-the-middle approach~\cite{horowitz1974computing}. An important open question in the theory of exact exponential time algorithms for hard problems~\cite{woeginger2008open} is whether any of these algorithms for \textsf{Subset Sum} can be improved. 

It has been shown in~\cite{jansen2016bounding,buhrman2015hardness} that under the Exponential Time Hypothesis, there is no $2^{o(n)} \cdot t^{o(1)}$ algorithm for \textsf{Subset Sum}. In another line of work, it has been shown that there is no $O(t^{1-\eps} \poly(n))$-time algorithm under the Set cover conjecture~\cite{cygan2016problems}, and also under $\seth$~\cite{abboud2022seth}. 

This still leaves open the question whether one can rule out the possibility of a $2^{cn} \cdot \poly(\log t)$-time algorithm for \textsf{Subset Sum} for some fixed constant $c < 1/2$ under a reasonable conjecture such as the Set Cover conjecture or $\seth$.

\subsection{Instance Compression} 

Harnik and Naor~\cite{HN10} studied  instance compression of $\NP$ decision problems. They focus on the problem with long instances but relatively small witness sizes, and ask whether the instance size can be reduced while preserving the information whether the input instance is in the language or not. Moreover, a compressed instance may not be of the same problem. This compressed instance can then be used to solve the problem in the future, maybe by using technological advances or with some algorithmic improvement.  A problem is said to be efficiently instance compressible if there exists a compression in size polynomial in the witness size, and polylogarithmic in the length of input size. They introduce new subclasses of $\NP$ depending on the compression size. They also study the implication of compression in cryptography. If we can have instance compression of problems like $\textsf{OR}$-$\SAT$, we can use these compressions to get cryptographic primitives, for example, collision-resistant hash functions. 

It is easy to construct an efficient compression of a $\textsf{GapSAT}$ instance when the gap is  $\left(1-\frac{1}{\poly(n)}\right)$. Create an instance of $\SAT$ by sampling a subset of clauses of $\poly(n)$ size from \textsf{GapSAT} instance uniformly at random. This instance is satisfiable if \textsf{GapSAT} instance is satisfiable; otherwise, it is unsatisfiable with high probability. Inspired by this compression, Harnik and Naor proposed the following problem: ``Can we have an efficient reduction from $\SAT$ to $\textsf{GapSAT}$ for which the number of variables in \textsf{GapSAT} depends only on the number of variables in the input SAT instance?" Note that a positive answer to this question will give a succinct PCP for the $\SAT$ problem. Later, Fortnow and Santhanam~\cite{FS11} gave a negative answer to the above question. More specifically, they showed that we could not have an instance compression for $\SAT$ problem unless $\coNP \subseteq \NPPoly$. They also point out that it will imply an impossibility of instance compression for Clique, Dominating set, and Integer Programming because of the known efficient reductions from $\SAT$ to these problems.

However, above mentioned results do not say anything about the compression of the $k$-$\SAT$ problem. It is easy to see that we can compress a $k$-$\SAT$ in $\widetilde{\mathcal{O}}(n^k)$ bits by removing the duplicate clauses. Dell and van Melkebeek~\cite{dell2014satisfiability} showed that this is almost the best possible compression we can hope to get. They showed that,  for any $\eps>0$, there is no compression algorithm that takes as input an arbitrary $k$-$\SAT$ instance, and outputs an equivalent instance of size $\mathcal{O}(n^{k-\eps})$ of some language $L'$, unless $\coNP \subseteq \NPPoly$. 

These barriers for instance compression of $\SAT$ only holds for probabilistic polynomial time reduction with \emph{no false negatives}. Drucker~\cite{drucker2015new} introduced the notion of \emph{probabilistic instance compression} which allows bounded errors on both side and showed that there is no non-trivial probabilistic instance compression for $OR$-$3$-$\sat$ unless there are non-uniform, statistical zero-knowledge proofs for all language in $\NP$.

\subsection{Our Results}
We focus on the fine-grained hardness of $\CVP$ in the $\ell_p$ norm for $p\in 2\intg^+$. We say that a probabilistic reduction does not have false negatives if, for any oracle call that is made with a \textsf{YES} instance, the oracle responds \textsf{YES} with probability $1$. We show the following impossibility result about reduction from satisfiability problem:

\begin{theorem}[Informal, see \cref{thm:SAT-to-CVP2,thm:SAT-to-CVPp}]\label{thm:inf_polytime}
For any even positive integer $p$ and constant $c>0$, there exists a constant $k_0$ such that for all $k > k_0$,  there is no polynomial time probabilistic reduction without false negatives from $k$-$\SAT$ on $n$-variables to $\CVP_p$ on $O(n^c)$ rank lattice that make at most $O(n^c)$ calls to $\CVP_p$ oracle, unless $\coNP$ is in $\NPPoly$.
\end{theorem}

\begin{figure}
    \centering
    \includegraphics[scale=0.75]{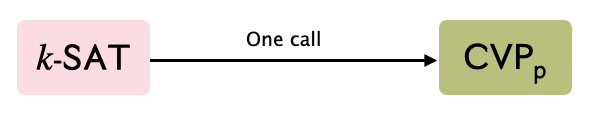}
    \caption{\cite{BGS17,ABGS19} gave a \emph{Karp} reduction from $k$-$\sat$ to $\CVP_p$ for $p\in [1,\infty]\setminus 2\intg$.}
    \label{fig:karp-reduction}
\end{figure}

\begin{figure}
    \centering
    \includegraphics[scale=0.75]{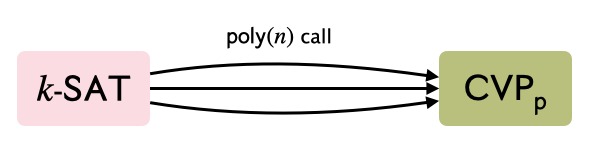}
    \caption{[This work] For $p\in 2\intg^+$, it is impossible to get a polynomial-time Turing reduction from $k$-$\SAT$ on $n$ variables to $\CVP_p$ on $n^c$ rank lattice unless $\coNP\subset \NPPoly$. }
    \label{fig:poly-time}
\end{figure}

\begin{figure}
    \centering
    \includegraphics[scale=0.75]{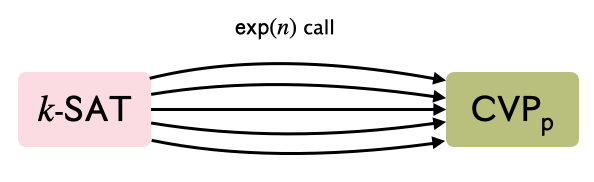}
    \caption{\cite{BGS17} gave $exp(n)$ time reduction from $k$-$\sat$ on $n$ variables to $\CVP_p$ instance of $O(n)$ rank lattice. It implies $2^{o(n)}$-hardness of $\CVP_p$ under \textsf{ETH}. It is an open problem to show $2^{cn}$-hardness of $\CVP_p$ for even $p$ under \textsf{SETH}.}
    \label{fig:exp-time}
\end{figure}

 For even $p$ greater than $2$, we need an additional promise on $\CVP_p$ instance that the target's distance from the lattice is at most $\exp(n^{O(1)})$ factor large than the shortest non-zero vector in the lattice. For $\SVP_p$ for all even $p$, we show the barrier for exact-$\SVP_p$(without any additional promise needed).

Our result says that for any even integer $p$ and constant $c>0$, it is not possible to get $2^{cn}$ $\seth$-hardness by a deterministic \emph{polynomial time} Karp reduction from $k$-$\SAT$ to $\CVP_p$ unless the polynomial hierarchy collapses to the third level. We also rule out Turing reductions, which makes less than $n^k$ calls to $\CVP_p$ oracle. These impossibility results also hold for a probabilistic reduction, as long as the oracle does not output false negatives. This result, in particular, explains why~\cite{BGS17, ABGS19} could not prove $\SETH$-hardness of $\CVP_p$ for even $p$. They showed $\seth$-hardness of $\CVP_p$ for non-even $p$ by a Karp-reduction (shown in \cref{fig:karp-reduction}) from $k$-$\SAT$ on $n$-variables to $\CVP_p$ on lattice of rank $n$. The above theorem says (under a complexity-theoretic assumption) that there exists a constant $k$ for which it is impossible to get a polynomial time-reduction Karp reduction from $k$-$\SAT$ to $\CVP_p$ on $O(n)$ rank lattice for even $p$. On the contrary, our result does not rule out all fine-grained polynomial-time reductions. For example, it does not mention the possibility of Turing reductions, which make $n^k$ calls to $\CVP_p$ oracle. Note that there also exists a polynomial time reduction from $k$-$\SAT$ to $\CVP_p$ on the lattice of rank $n^{O(k)}$, but this reduction does not say anything about $2^{cn}$ fine-grained hardness of $\CVP$.    

Notice that the above theorem does not say anything about the possibility of a super-polynomial time reduction. We know that $\CVP_p$ for any $p\geq 1$ is $2^{o(n)}$-hard (shown in \cref{fig:exp-time}) assuming the Exponential Time Hypothesis~\cite{BGS17}. This reduction from $k$-$\SAT$ makes exponential number of calls to the $\CVP_p$ oracle.

If we believe that $\coNP$ is not contained in $\NP/\poly$, then we can only hope to get $\SETH$-hardness for $\CVP_p$ for even $p$ by a super-polynomial time reduction. Moreover, we  give a barrier for a specific class of super-polynomial time reductions. We  generalize our result for a range of different running times.

\begin{theorem}[Informal, see \cref{cor:exptime-SAT-to-CVP,cor:exptime-SAT-to-CVPp}]\label{thm:inf_exptime}
    For any even positive integer $p$ and constant $c>0$, there exists a constant $k_0$ such that for all $k > k_0$ and $T$,  there is no $T$-time probabilistic reduction with no false negatives from $k$-$\SAT$ on $n$-variables to $\CVP_p$ on $O(n^c)$ rank lattice that make at most $O(n^c)$ calls to a $\CVP_p$ oracle, unless $\coNP$ is in $\frac{\mathsf{NTIME}(\poly(n)\cdot T)}{\mathsf{Poly}}$.
\end{theorem}

Notice that the above result shows a barrier for super-polynomial time-reductions that make only polynomial number of calls to the $\CVP_p$ oracle. We also study the barriers for probabilistic polynomial time reductions. However, we are only able to show this barrier for \textit{a non-adaptive} reduction from $k$-$\sat$ to $\CVP_p$ for even $p$.

\begin{theorem}[Informal, see \cref{thm:rand-poly-time-compress-CVP2,cor:rand-poly-time-compress-CVPp}]\label{thm:inf_polytime_prob}
For any even positive integer $p$ and constant $c>0$, there exists a constant $k_0$ such that for all $k > k_0$,  there is no polynomial time probabilistic non-adaptive reduction from $k$-$\SAT$ on $n$-variables to $\CVP_p$ on $O(n^c)$ rank lattice that make at most $O(n^c)$ calls to the $\CVP_p$ oracle, unless there are non-uniform, statistical zero-knowledge proofs for all languages in $\NP$.
\end{theorem}

We also observe that we can conclude the (im)possibility of $\SETH$-hardness of \textsf{Subset-Sum} to get the following result.

\begin{theorem}[Informal, see \cref{thm:barrier-for-subset-sum}]
For any constant $c>0$, there exists a constant $k_0$ such that for all $k > k_0$,  there is no polynomial time probabilistic reduction without false negatives from $k$-$\SAT$ on $n$-variables to \textsf{Subset-Sum} on $O(n^c)$ numbers that make at most $O(n^c)$ calls to \textsf{Subset-Sum} oracle, unless $\coNP$ is in $\NPPoly$.
\end{theorem}

\subsection{Our Techniques}
Instance compression of computational problems has interesting connection in \emph{fixed-parameter-tractable} algorithms~\cite{downey2012parameterized,guo2007invitation} and Cryptography~\cite{HN10}. In this work, we initiate the study of instance compression for lattice problems, and shows its consequences to $\seth$ hardness. 

\paragraph{Instance compression and Fine-grained hardness:} We show a connection between the $\SETH$-hardness and Instance compression. Let's say we are interested in fine-grained $\SETH$-hardness of problem \textsf{A} for which there exists a polynomial time algorithm that gives instance compression of polynomial size. Note that, to show $\seth$-hardness we need efficient reduction from $k$-$\sat$ for all constant $k$. If there exists a fine-grained polynomial time reduction from $k$-$\SAT$ to problem \textsf{A} then it will immediately give an algorithm for polynomial size compression (independent of $k$) of $k$-$\SAT$. However, Dell and van Melkebeek~\cite{dell2014satisfiability} showed that there does not exist any non-trivial compression for $k$-$\SAT$ problems unless the polynomial hierarchy collapses to the third level. In other words, if a computational problem has a  polynomial size compression then it is impossible to get $\seth$-hardness by a polynomial time reduction unless the polynomial hierarchy collapses to the third level. Moreover, \cite{dell2014satisfiability} gives the barrier for non-trivial oracle communication protocol for $k$-$\sat$. Oracle communication protocol (\cref{def:oracle-communication-protocol}) can be seen as a generalized notion of instance compression. Using the impossibility of oracle communication protocol, we show a barrier for \emph{adaptive} fine-grained reductions from $k$-$\SAT$ to a problem with polynomial size compression.

We also give barriers for $\seth$ hardness of instance compressible problem by bounded error polynomial time \emph{probabilistic} reduction. Drucker~\cite{drucker2015new} showed that it is impossible to get a non-trivial probabilistic instance compression of $k$-$\sat$ unless there are non-uniform, statistical zero-knowledge proofs for all languages in $\NP$. Using a similar argument as above, we show a barrier for polynomial time probabilistic reduction from $k$-$\sat$ to a problem that has a probabilistic algorithm for instance compression. However, we don't know any barrier for the probabilistic oracle communication protocol, so we can only show a barrier for probabilistic non-adaptive fine-grained reduction.

Moreover, our work suggests that all computational problems can be classified into two classes: (i) problems that have a fixed polynomial size instance compression and (ii) problems for which it is not possible to find such compression. It is impossible to get polynomial-time fine-grained reductions from problems in (ii) class to problems in (i) by a polynomial time reduction.

\paragraph{Instance compression of $\CVP$ in even norm:}
We present a polynomial time algorithm that gives $\mathcal{O}(n^{p+3})$ bit-length instance compression for almost exact $\CVP_p$ for even $p$ on rank $n$ lattice. For this purpose, we introduce variants of $\CVP_p$; $\CVPip$ and $\CVPmvp$. In $\CVPip$, given the inner product of basis vectors and target vector, the goal is to find the coefficient of a closest lattice vector to the target. This problem is well-defined from the fact that the euclidean distance between any lattice point and target vector can be computed, given the inner products of basis vectors and target, and coefficient of the lattice point in the underlying basis. So, there is also a trivial reduction from $\CVP_2$ to $\CVPip$. $\CVPmvp$ is an extension of the $\CVPip$ for $\ell_p$ norms when $p$ is even. We show a polynomial time algorithm that reduces arbitrary $\CVP_p$ instance for even $p$ to $\CVPip$/$\CVPmvp_p$ instance of fixed polynomial size. 

For simplicity, here we will only present a sketch of compression algorithm for Euclidean norm. As mentioned above any $\CVP_2$ instance can be compressed by just storing the $(n+1)^2$ pairwise (with repetitions) inner products of the basis vectors $\vect{b}_1, \ldots, \vect{b}_n$, and the target vector $\vect{t}$. The instance compression for $\CVP$ would be immediate from this if all co-ordinates of the vectors are bounded by $2^{n^c}$ for some constant $c$, since that would imply a compression of (exact) $\CVP$ to an instance of $\CVP^{\IP}$ of size $(n+1)^2\log \left(m\cdot 2^{2n^c}\right)=(n+1)^2(2n^c+\log m)$. In the following, we show how to decrease all the co-ordinates of the $\CVP$ instance while still retaining information of whether the instance is a YES/NO. For this, we need to consider approximate-$\CVP$ (with an approximation factor very close to $1$) rather than exact $\CVP$. 

First, we transform the basis and target vector such that the coefficients of the closest lattice vector to the target vector are bounded by $2^{n^2}$. To do this, we use the \textsf{LLL} algorithm~\cite{LLL82}. Next, we want to bound the distance from the closest lattice vector by $2^{n^2}$. To achieve this, we divide the lattice basis vectors and target vector by a carefully chosen large integer and ignore the fractional part of the vectors. As long as we only allow lattice vectors with coefficients bounded by $2^{n^2}$, we show that this truncation of lower-order bits does not introduce any other close vectors. This implies that $\CVP$ reduces to a variant of $\CVP$ where the goal is to find a closest lattice vector whose coefficient is bounded by $2^{n^2}$, and the distance from target is also bounded by $2^{n^2}$. This step essentially uses the (small) gap between the \textsf{Yes} and \textsf{NO} instance\footnote{ Note that if the distance from target vector is bounded by $2^{n^2}$ then we get instance compression for exact-$\CVP_2$}. Then, we reduce this to an instance where the basis and target vector coordinates are bounded by $2^{n^2}$. For this, we choose a prime number significantly larger than the distance of the target from closest lattice vector and reduce all coordinates of the basis and target vector modulo the prime. The randomness of the prime is sufficient to guarantee that, with high probability, we do not introduce any new close vectors with bounded coefficients. So, finally, we reduce $\CVP$ to a variant of $\CVP$ with all coordinates bounded by $O(n^2)$. More specifically, we reduce it to the problem of finding closest lattice vector with bounded coefficients. Now we reduce it to $\CVPip$ by computing the inner products of basis vectors and target.

We also demonstrate an another technique for instance compression for $\CVP_2$ by utilizing a theorem from \cite{FrankT87}. First, we apply a trivial reduction from $\CVP_2$ to $\CVPip$. Then, we employ a result from the theorem in \cite{FrankT87} as a black-box to reduce it into a $\CVPip$ instance with inner products bounded by $2^{O(n^2)}$. It's worth noting that this instance compression technique is applicable to exact-$\CVP_2$. Furthermore, this technique can be extended to any even norm but requires an additional promise: that the distance of the target from the lattice is at most $\exp(n^{c})$ times the length of the shortest non-zero lattice vector.

\subsection{Comparison to previous works}
In literature, there are  results~\cite{carmosino2016nondeterministic, belova2023polynomial,ABB+LatticeProblemsPolynomial2022} that show the barrier for getting \textsf{SETH}-hardness of problems. In \cite{carmosino2016nondeterministic},  authors propose Non-deterministic Strong Exponential Time Hypothesis(\textsf{NSETH}), which states that for every $\eps>0$ there exists a $k$ so that $k$-taut is not in $\textsf{NTIME}(2^{n(1-\eps)})$, where $k$-taut is the language of all $k$-DNF which are tautologies. They gave faster co-nondeterministic algorithms for 3-SUM, APSP and model checking of a large class of first-order graph properties. They show that it is unlikely to get a fine-grained deterministic reduction from $k$-$\SAT$ to these problems. If there is a fine-grained reduction then it implies that $k$-taut has faster non-deterministic algorithm which contradicts \textsf{NSETH}.

In \cite{belova2023polynomial}, the authors investigate the barriers to proving the \textsf{SETH}-hardness of Hamiltonian Path, Graph Coloring, Set Cover, Independent Set, Clique, Vertex Cover, and 3d-Matching. Specifically, they show that if a fine-grained reduction exists from $k$-$\SAT$ to any of these problems, it would imply new circuit lower bounds. In comparison to these results, our work focuses on ruling out fine-grained reductions for lattice problems and Subset-Sum under weaker conditions than those used in previous techniques. However, it should be noted that our conclusion is relatively weaker, as we cannot rule out Fine-grained Turing reductions that make superpolynomial calls.

Recently, in \cite{ABB+LatticeProblemsPolynomial2022}, authors shows barriers for $\SETH$-hardness of constant approximation of $\CVP$. This result does not say anything about $\SETH$-hardness of near exact-$\CVP$.

\subsection{Other Conclusion and Open Questions.}

There are several interesting observations that can be made about our main result in light of prior work. In the following, let $q$ be quantified over $[1,\infty) \setminus 2\mathbb{Z}$, and $p$ be quantified over $2\mathbb{Z}^+$.

\begin{itemize}
    \item It was shown in~\cite{ABGS19} that for all $q$, $(1+\eps)$-approximate $k$-$\SAT$ on $n$ variables can be reduced to $(1+\eps/\poly(k))$-$\CVP_q$ on a rank $n$ lattice. Notice that without loss of generality, one may assume that the number of clauses of a $k$-SAT instance is at most $O(n^k)$, and thus $(1+1/n^k)$-approximate $k$-$\SAT$ is the same as $k$-$\SAT$. This implies that, from our result, one can conclude that there does not exist a $\poly(n)$-time reduction from $(1+1/\poly(n))$-$\CVP_q$ on a rank $n$ lattice to $\CVP_2$ on a $\poly(n)$-rank lattice for any $q \in [1,\infty) \setminus 2\mathbb{Z}$. Our result provides evidence that shows that $\CVP_2$ might be easier than $\CVP_q$ for $q \in [1,\infty) \setminus 2\mathbb{Z}$. This conclusion can also be made with $\CVP_2$ replaced by $\CVP_p$, with a mild caveat that the $\CVP_p$ instance must satisfy the promise that the distance of the target from the lattice is at most an $\exp(n^{O(1)})$ factor larger than the shortest non-zero vector in the lattice. 
    \item This result should be contrasted with~\cite{RR06} which showed that, approximate $\CVP_2$ is reducible to approximate $\CVP_p$ with almost the same approximation factor, which also gave evidence that $\CVP_2$ might be easier than $\CVP_p$ in other $\ell_p$ norms.

\end{itemize}

Our work helps take further our understanding of the limitations of the fine-grained hardness of $\CVP$ and $\SVP$ under  the Strong Exponential Time Hypothesis. Some of the questions left open by our work are as follows.

\begin{itemize}

\item One interesting question that emerges from our work is the following. Prior work has shown that it is much easier to make algorithmic progress for $\CVP,\SVP$ in the Euclidean norm~\cite{MV13,ADRS15,ADS15}, as opposed to the corresponding problems in other $\ell_p$ norms. To our understanding, this was partially because our understanding of the Euclidean norm is much better than that for other norms. This work suggests that perhaps computational problems in the $\ell_2$ norm (and other even norms) are inherently easier, which suggests trying to find faster algorithms for $\SVP,\CVP$ in $\ell_p$ norms where $p$ is an even integer. 

\item We need to introduce an additional promise on the  $\CVP_p$ instance for $p > 2$ for our compression algorithm to work. This doesn't seem inherent, and is likely just a consequence of our techniques. The problem of removing this restriction is left open.

\item While our work rules out the possibility of $\poly(n)$-time reduction from $k$-$\SAT$ to $\CVP_p$ and to the \textsf{Subset-Sum} problem, we do not rule out the possibility of such a reduction that makes more than $\poly(n)$ calls to the oracle for the respective problems. Ruling out such a reduction under a reasonable conjecture is a very interesting open question. 
\end{itemize}

\paragraph{Organization:} We give the preliminaries in \cref{sec:prelims}. In \cref{sec:variants-of-CVP}, we propose the problems $\CVPip$ and $\CVPmvp$. In \cref{sec:compress_cvp_euclidean_norm}, we give an instance compression algorithm for $\CVP$ in the Euclidean norm. We extend this and give an instance compression algorithm for $\CVP_p$ for even $p$ in \cref{sec:compress_cvp_even_norm}. In \cref{sec:compresssion_exact}, we present compression algorithm for exact-CVP in even norm. These results are significantly better than previous two sections but uses a theorem from \cite{FrankT87} as black-box. We present the barriers for $\seth$-hardness of $\CVP$ in $\ell_p$ norms for even $p$ in \cref{sec:barriers}.  We show barrier for $\seth$ hardness of \textsf{Subset-Sum} problem  in \cref{sec:subsetsum}.
\section{Preliminaries}
\label{sec:prelims}

We use the notation $\real$, $\ratn$ and $\intg$ to denote the set of real numbers, rational numbers and integers respectively. For any integer $n>0$, we use $[n]$ to denote the set $\{1,\ldots,n\}$. For any integers $a,b(>a+1)$, we use $[a,b]$ to denote the set $\{a+1,\ldots,b-1\}$. We will use the boldfaced letters (for example $\vect{x}$) to denote a vector and, denote $\vect{x}$'s coordinate by $x_i$ indices. We use bold capital letters (for example $\mathbf{B}$) to denote a matrix. We will use the notation $\lfloor a \rceil $ to denote the closest integer to $a$, and $\lfloor a \rfloor$ to denote the greatest integer less than equal to $a$. For any vector $\vect{v}$, we will use the notation $\lfloor \vect{v}\rfloor$ to denote the vector representing the floor of each coordinate of the vector.  For any $p\in [1,\infty)$, the $\ell_p$ norm on $\real^m$ is defined as follows:
\[\|\vect{x}\|_p:=\left(\sum\limits_{i=1}^{m}|x_i|^p\right)^{1/p},\]
and $\ell_\infty$ norm is defined as 
\[\|\vect{x}\|_\infty =\max\{|x_i|\}.\]
We will use the following inequality between different $\ell_p$ norms,
\[\text{for any } p\leq q, \; \vect{x} \in \real^m, \; \|\vect{x}\|_q\leq \|\vect{x}\|_p \leq m^{\frac{1}{p}-\frac{1}{q}}\|\vect{x}\|_q .\]
We will usually drop the subscript and  use $\|\vect{x}\|$ to denote $\|\vect{x}\|_2$. We often shorthand $p \in [1,\infty) \cup \{\infty\}$ by $p \in [1,\infty]$.

For any $\vect{v}_1, \ldots, \vect{v}_i \in \real^m$, we denote by $\proj_{\{\vect{v}_1,\cdots, \vect{v}_{i-1}\}^{\perp}}\vect{v}_i$, the vector formed by projecting $\vect{v}_i$ orthogonal to the subspace spanned by $\vect{v}_1,\cdots, \vect{v}_{i-1}$.

\paragraph{Lattices:} Let
$\basis=\{\vect{b_1},\cdots,\vect{b}_n\}$ be a set of $n$ linearly independent vectors from $\real^m$ for some positive integers $m, n$ with $m \ge n$. The lattice $\cL$ generated by basis $\basis$ is defined as follows:
\[\cL(\basis):=\left\{ \sum\limits_{i=1}^n z_i\vect{b}_i\; :\; z_i\in \intg \right\}.\]
Here $n$ is called the rank of the lattice and, $m$ is called the dimension of the lattice. Note that a lattice has infinitely many bases. We often write the basis $\basis$ as a matrix in $\real^{m \times n}$. For any $p\in [1,\infty]$, we use $\lambda_1^{(p)}(\cL)$ to denote the length of a shortest non-zero vector in $\ell_p$ norm, 
\[\lambda_1^{(p)}(\cL):=\min\left\{\|\vect{v}\|_p:\vect{v}\in \cL \text{ and } \vect{v}\neq \vect{0}\right\}.\]
For any vector $\vect{t} \in \real^m$, we use $\dist_p(\cL,\vect{t})$ to denote the distance in $\ell_p$ norm of the target vector  from the lattice $\cL$, 
\[\dist_p(\cL,\vect{t}):= \min\left\{\|\vect{v}-\vect{t}\|_p\; : \; \vect{v}\in \cL \right\}.\]

For the purpose of computational problems (and hence for the rest of the paper), we restrict our attention to lattices with basis entries in $\ratn$.

\subsection{Lattice Problems}

In the following, we introduce lattice problems that we study in this paper.

\begin{defn}[$\gamma$-$Gap\CVP_p$]
For any $\gamma = \gamma(m,n) \geq 1$ and $p\in [1,\infty]$, the $\gamma$-$Gap\CVP_p$ (Closest Vector Problem) is the decision problem  defined as follows: Given a basis $\basis\in \ratn^{m\times n}$ of lattice $\cL$, a target vector $\vect{t}\in \ratn^m$ and a number $d>0$, the goal is to distinguish between a \textsf{YES} instance, where $\dist_p(\cL,\vect{t})\leq d$ and a \textsf{NO} instance, where $\dist_p(\cL,\vect{t})>\gamma d$.
\end{defn}

\begin{defn}[$\gamma$-$Gap\CVP_p^{\phi}$]
For any $\gamma = \gamma(m,n) \geq 1$, $p\in [1,\infty]$ and a positive real-valued function $\phi$ on lattice, the $\gamma$-$Gap\CVP_p^{\phi}$ (Closest Vector Problem) is the decision problem  defined as follows: Given a basis $\basis\in \ratn^{m\times n}$ of lattice $\cL$, a target vector $\vect{t}\in \ratn^m$ and a number $d>0$ with the promise that $\dist_p(\cL,\vect{t})\leq \phi(\cL)$, the goal is to distinguish between a \textsf{YES} instance, where $\dist_p(\cL,\vect{t})\leq d$ and a \textsf{NO} instance, where $\dist_p(\cL,\vect{t})>\gamma d$.
\end{defn}

\begin{defn}[$\gamma$-$Gap\SVP_p$]
For any $\gamma = \gamma(m,n) \geq 1$ and $p\in [1,\infty]$, the $\gamma$-$Gap\SVP_p$ (Shortest Vector Problem) is the decision problem  defined as follows: Given a basis $\basis\in \ratn^{m\times n}$ of lattice $\cL$, and a number $d>0$, the goal is to distinguish between a \textsf{YES} instance, where $\lambda_1^{(p)}(\cL)\leq d$ and a \textsf{NO} instance, where $\lambda_1^{(p)}(\cL)>\gamma d$.
\end{defn}

We omit the parameter $\gamma$ if $\gamma=1$ and the parameter $p$ if $p=2$.

\subsection{LLL Algorithm}
For any set of vectors $\basis=\{\vect{b}_1,\cdots, \vect{b}_n\}\in \ratn^{m\times n}$, we define the Gram Schmidt Orthogonalization (GSO) of $\basis$ as $\vect{b}_1^*,\cdots,\vect{b}_n^*$, where 
\[\vect{b}_i^*:=\proj_{\{\vect{b}_1^*,\cdots, \vect{b}_{i-1}^*\}^{\perp}}\vect{b}_i.\]
and the Gram Schmidt coefficients as
\[\mu_{ij}:=\frac{\langle \vect{b}_i, \vect{b}_j^*\rangle }{\|\vect{b}_j^*\|^2}.\]
Here $\{\vect{b}_1^*,\cdots,\vect{b}_{i-1}^*\}^\perp$ denotes the subspace of $\real^m$ which is orthogonal to space formed by $\vect{b}_1^*,\cdots,\vect{b}_{i-1}^*$.
\begin{theorem}[\cite{LLL82}]\label{thm:lll}
For any positive integers $m$ and $n$, there exists an algorithm that given a basis $\mathbf{C}=\{\vect{c}_1,\cdots,\vect{c}_n\}\in \ratn^{m\times n}$  of lattice $\cL$, outputs a basis $\basis=\{\vect{b}_1,\cdots, \vect{b}_n\}\in \ratn^{m\times n}$ of lattice $\cL$ whose GSO vectors and coefficients satisfy the following conditions:
 \begin{enumerate}
     \item $\forall \; n\geq i>j\geq 1$, $|\mu_{ij}|\leq \frac{1}{2}$.
     \item $\forall \; n \geq i > 1$, $\|\vect{b}_i^*\|^2\geq \frac{1}{2} \| \vect{b}_{i-1}^*\|^2$.
     \item $\|\vect{b}_1\|\leq 2^{n/2}\cdot \lambda_1(\cL)$. 
 \end{enumerate}
 We call the output basis as an \textsf{LLL}-reduced basis. The algorithm runs in time polynomial in the size of the input basis.
 \end{theorem}

 \subsection{\textsf{k-SAT} and \textsf{Subset Sum}}
 
\begin{defn}[$k$-$\sat$]
For any positive integers $k,n,m$, given a \textsf{CNF} formula $\Psi$ of $m$ clauses over $n$ Boolean variables and each clause of $\Psi$ contains at most $k$ literals, the goal is to distinguish between a \textsf{YES} instance, where there exists an assignment that satisfies $\Psi$ and a \textsf{NO} instance, where does not exist any satisfying assignment.
\end{defn}

Impagliazzo and Paturi~\cite{IP01} gave the following two hypotheses about the hardness of $k$-$\sat$. These two hypotheses are widely used to prove the fine-grained hardness of many computational problems.
 
 \begin{defn}[Exponential Time Hypothesis (\textsf{ETH})]
The Exponential Time Hypothesis says the following: for any $k\geq 3$ there exists a constant $\eps >0$ such that $k$-$\sat$ can not be solved in $2^{\eps n}$ time. In particular, $3$-$\sat$ can not be solved in $2^{o(n)}$ time.  
 \end{defn}

 \begin{defn}[Strong Exponential Time Hypothesis (\textsf{SETH})]
The Strong Exponential Time Hypothesis says the following: for any constant $\eps>0$, there exists a constant $k$ such that $k$-$\sat$ can not be solved in $2^{(1-\eps) n}$ time.  
\end{defn}

We also define the \textsf{Subset-Sum} problem.

\begin{defn}[\textsf{Subset-Sum}]
For any positive integer $n>0$, given a set of integers $\cS=\{a_1,\cdots, a_n\}$ and target $t$, the goal is to decide whether there exists a subset of $S$, whose elements sums to $t$.
\end{defn}

 \subsection{Instance Compression}

In this work, we will use two variants of instance compression, probabilistic instance compression with \emph{no false negative} and probabilistic instance compression that allows both side error. We will use instance compression for probabilistic instance compression with no false negative.

\begin{defn}[Instance Compression]
A decision problem $P$ is $(f,g,\xi)$ instance compressible with soundness error bound $\xi$ if there exists an $f$-time randomized reduction from any arbitrary instance $x$ of $P$ to some instance $x'$ of size $g$ of decision problem $P'$ such that,  if $x$ is a \textsf{YES} instance of $P$ then $x'$ is a \textsf{YES} instance of $P'$, and if $x$ is a \textsf{NO} instance of $P$ then $x'$ is a \textsf{NO} instance of $P'$ with probability at least $1-\xi$. Here, $f$ and $g$ can be a functions of the witness size and the bit length of $x$. 
\end{defn}

\begin{defn}[Probabilistic Instance Compression]
A decision problem $P$ is $(f,g,\xi)$ probabilistic instance compressible with error bound $\xi$ if there exists an $f$-time randomized reduction from any arbitrary instance $x$ of $P$ to some instance $x'$ of size $g$ of decision problem $P'$ such that, if $x$ is a \textsf{YES} instance of $P$ then $x'$ is a \textsf{YES} instance of $P'$ with probability at least $1-\xi$, and if $x$ is a \textsf{NO} instance of $P$ then $x'$ is a \textsf{NO} instance of $P'$ with probability at least $1-\xi$. Here, $f$ and $g$ can be a functions of the witness size and the bit length of $x$. 
\end{defn}

We use the definition of Oracle Communication protocol from \cite{dell2014satisfiability}. It can also be seen as a generalization of instance compression.
\begin{defn}[Oracle Communication protocol]\label{def:oracle-communication-protocol}
An $(f,g)$ oracle communication protocol for a language $L$ is a communication protocol between two players. The first player is given the input $x$ and has to run in time $f$: the second player is computationally unbounded but not given any part of $x$. At the end of the protocol, the first player outputs \textsf{YES} or \textsf{NO}. It always outputs $\textsf{YES}$ if $x\in L$ and outputs \textsf{NO} with probability atleast constant if $x\not\in L$. The cost of the protocol $g$ is the number of bits of communication from the first player to the second player. Again, $f, g$ can be a functions of the witness size and the bit length of $x$. The first player is allowed to use randomness but the output by the first player is assumed to be a deterministic function of the communication transcript.\footnote{We include randomness in our definition of the oracle communication protocol, since our instance compression algorithms require randomness.}
\end{defn}
 
 We will use the following theorem given by the Dell and van Melkebeek~\cite{dell2014satisfiability}  about the sparsification of the Satisfiability problem.

\begin{theorem}\label{thm:compress-SAT}
Let $k\geq 3$ and $\epsilon>0$ a positive real. There is no oracle communication protocol for $k$-$\sat$ of cost $\mathcal{O}(n^{k-\eps})$ that runs in time $\poly(n)$, unless $\coNP \subseteq \NP/\poly$.
\end{theorem}

 Our definition of Oracle communication protocol allows the first player to do randomized operations. The Dell and van Melkebeek's proof still holds for this generalized definition.
 We generalize the impossibility of oracle communication protocol for $k$-$\SAT$ within any time $T$.
 
 \begin{restatable}{theorem}{GenOracleCom}\label{thm:randomized-oneside-compress}
Let $k\geq 3$ and $\epsilon>0$ a positive real. For any $T=T(n)>0$, there is no oracle communication protocol  for $k$-$\sat$,  of cost $\mathcal{O}(n^{k-\eps})$ that runs in time $T$, unless $\coNP \subseteq \cc{NTIME(\poly(n)\cdot T)}/\poly$.
\end{restatable}
  We defer the proof of generalized theorem to Section~\ref{sec:compression-impossibility}.

It is considered unlikely that $\coNP \subseteq \NPPoly$, it also implies that polynomial hierarchy collapses to third level.

\begin{theorem}[\cite{yap1983some}]
    If $\coNP \subseteq \NPPoly$, then polynomial hierarchy collapses to third level.
\end{theorem}

We will require the following results from \cite{dell2014satisfiability}.

\begin{lemma}\cite{dell2014satisfiability}[Lemma~2]\label{lem:OR-SAT-Clique}
For any integer $k\geq 2$, there is a polynomial time reduction from OR($3$-$\sat$) to $k$-Clique that maps $t$ tuples of instances of bitlength $n$ each to an instance of $O(n\cdot \max(n,t^{1/k+o(1)}))$ vertices.
\end{lemma}

\begin{lemma}\cite{dell2014satisfiability}[Lemma~5]\label{lem:VCover-SAT}
For any $k\geq 3$, there is a polynomial time reduction from $k$-Vertex Cover to $k$-$\sat$ that maps a $k$-uniform hypergraph on $n$ vertices to $k$-CNF formula on $O(n)$ variables.
\end{lemma}

For the definitions of Hypergraph problems, $k$-Vertex Cover and $k$-Clique, we refer the reader to the preliminaries section of \cite{dell2014satisfiability}. Note that there is a trivial reduction from $k$-Clique to $k$-Vertex Cover, which also preserves the number of vertices.

\section{{CVP inner product}, {CVP multi vector product}: variants of {CVP}}
\label{sec:variants-of-CVP}

We introduce the following new variant of the problems,   $\gamma$-$Gap\CVP^{\mathsf{IP}}$, where the input lattice and the target vectors are not given directly but as pairwise inner products. More precisely, the input is $\alpha_{i,j} \in \mathbb{Q}$ for $1 \le  j < i \le n$, and $\beta_i \in \mathbb{Q}$ for $1 \le i \le n+1$ such that the corresponding lattice and the target vector satisfies the following. For $1 \le j \le i \le n$
\[
\alpha_{i,j} = \langle \vect{b}_i, \vect{b}_j \rangle \;,
\]
\[
\beta_{i} = \langle \vect{b}_i, \vect{t} \rangle \;,
\]
and 
\[
\beta_{n+1} = \langle \vect{t}, \vect{t} \rangle \;. 
\]
Note that the square of the $\ell_2$ distance between any integer combinations of the basis vectors and the target is an integer combination of $\alpha_{i,j}$'s and $\beta_i$'s.

For any $p\in 2\intg^{+}$, we extend the notion of inner product to multi vector product ($\mathsf{mvp}$)  defined as follows:
\[\forall \vect{v}_1,\vect{v}_2,\cdots,\vect{v}_p\in \real^m,\; \mathsf{mvp}(\vect{v}_1,\vect{v}_2,\cdots,\vect{v}_p):=\sum_{i=1}^m \left(\prod_{j=1}^p v_{ij}\right)\]
We call these variants of the problems  as $\gamma$-$\GapCVP_p^{\mathsf{mvp}}$ where the input lattice and the target vector  are not given directly but as $(n+1)^p$ multi vector products. Let $\vect{b}_{n+1}=\vect{t}$. Then, the input is $\alpha_{i_1,i_2,\cdots,i_p}$ for $1 \le  i_1 \le i_2\cdots \le i_p \le n+1$, such that the corresponding lattice and the target vector satisfies the following. For $1 \le  i_1 \le i_2\cdots \le i_p \le n+1$
\[
\alpha_{i_1,i_2,\cdots,i_p} = \mathsf{mvp}(\vect{b}_{i_1},\vect{b}_{i_2},\cdots,\vect{b}_{i_p}) \;.
\]

\begin{lemma}\label{lem:cvp-mvp}
For any $p\in 2\intg^+$, and vectors $\vect{v}_1,\cdots,\vect{v}_n$, for any $a_1,\cdots,a_n\in \intg$, $\|a_1\vect{v}_1+\cdots +a_n\vect{v}_n\|_p^p$ can be computed in polynomial time given only $a_1,\cdots,a_n$, and  $\mathsf{mvp}(\vect{v}_{i_1},\cdots,\vect{v}_{i_p} )$ for all $ i_1,\cdots,i_p\in [n]$.
\end{lemma}

\begin{proof}
From the definition of the $\ell_p$ norm, we get 
\begin{align*}
    \|a_1&\vect{v}_1+\cdots+a_n\vect{v}_n\|_p^p\\
    &=\sum_{i=1}^m |a_1v_{1i}+\cdots + a_n v_{ni}|^p & (\text{by }\ell_p \textsf{ norm definition})\\
    &=\sum_{i=1}^m (a_1v_{1i}+\cdots + a_n v_{ni})^p & (\textsf{because } p \textsf{ is a positive even integer} )\\
    &=\sum_{i=1}^m \sum_{(j_1,\cdots,j_p)\in [n]^p} (a_{j_1}v_{j_1 i})\cdot (a_{j_2}v_{j_2 i}) \cdots (a_{j_p} v_{j_p i}) & \\
    &=\sum_{i=1}^m \sum_{(j_1,\cdots,j_p)\in [n]^p} (a_{j_1} \cdot a_{j_2} \cdots a_{j_p}) (v_{j_1 i} \cdot v_{j_2 i}\cdots   v_{j_p i}) & \\
    &=\sum_{(j_1,\cdots,j_p)\in [n]^p} (a_{j_1} \cdot a_{j_2} \cdots a_{j_p}) \sum_{i=1}^m  (v_{j_1 i} \cdot v_{j_2 i}\cdots   v_{j_p i})& \\
    &=\sum_{(j_1,\cdots,j_p)\in [n]^p} (a_{j_1} \cdot a_{j_2} \cdots a_{j_p})\cdot \mathsf{mvp}(\vect{v}_{j_1},\cdots,\vect{v}_{j_p}) & \textsf{ (by definition of mvp)}
\end{align*}

The lemma follows from the above equation.
\end{proof}
Notice that the above lemma implies that for even $p$, $p^{\textsf{th}}$ power of the $\ell_p$ distance between any integer combinations of the basis vectors and the target is an integer linear combination of the $(n+1)^p$ integers $\alpha_{i_1,i_2,\cdots,i_p}$'s. Hence, for even $p$, we can efficiently reduce $\gamma$-$\GapCVP_p$ to $\gamma$-$\GapCVPmvp_p$. 

We define variants of $\CVPmvp_p$, $(r,q)$-$\CVPmvp_p$ and $r$-$\CVPmvp_p$.
\begin{defn}[$(r,q)$-$\CVPmvp_p$]\label{def:r-q-cvpmvp}
    For any positive integers $q= q(m,n)$ and $r = r(m,n)$, the $(r,q)$-$\CVPmvp_p$ is the promise problem  defined as follows: Given $\textsf{mvp}$ form of a basis $\basis\in \ratn^{m\times n}$ and a target vector $\vect{t}\in \ratn^m$ and a number $d>0$, the goal is to distinguish between a `\textsf{YES}' instance, where $\exists \vect{z}\in [-r,r]^n$ $\|\basis\vect{z}-\vect{t}\|_p^p \mod q \leq d$ and a `\textsf{NO}' instance, where  $\forall \vect{z}\in [-r,r]^n$, $\|\basis\vect{z}-\vect{t}\|_p^p\mod q> d$.
\end{defn}
    
When $p=2$ we will also denote $(r,q)$-$\CVPmvp_p$ by $(r,q)$-$\CVPip$.

\begin{defn}[$r$-$\CVPmvp$]
    For any positive integer $r = r(m,n)$, the $(r)$-$\CVPmvp$ is the promise problem  defined as follows: Given \textsf{mvp} form of a basis $\basis\in \ratn^{m\times n}$ and a target vector $\vect{t}\in \ratn^m$ and a number $d>0$, the goal is to distinguish between a `\textsf{YES}' instance, where $\exists \vect{z}\in [-r,r]^n$ $\|\basis\vect{z}-\vect{t}\|_p^p \leq d$ and a `\textsf{NO}' instance, where  $\forall \vect{z}\in [-r,r]^n$, $\|\basis\vect{z}-\vect{t}\|_p^p> d$.
\end{defn}
When $p=2$ we will also denote $r$-$\CVPmvp_p$ by $r$-$\CVPip$.

\section{Instance compression for {almost exact-CVP} }\label{sec:compress_cvp_euclidean_norm}
In this section, we present an instance compression algorithm for $\textsf{CVP}$ in the Euclidean norm. We show that, for any $c > 0$, and for any $\eps\geq 2^{-n^c}$, we can reduce an instance of $(1+\eps)$-$\GapCVP$ with rank $n$, and dimension $m$ to an instance of $(r,q)$-$\CVPip$, such that the size of the $(r,q)$-$\CVPip$ instance is $n^{c_\eps}$ for some $c_\eps>0$. The reduction takes time polynomial in the input size.

In the following lemma, we show using Babai's algorithm~\cite{babai86} that without loss of generality, we may assume that given a CVP instance $(\mathbf{B}, \vect{t})$, the coefficient vector of the vector closest to $\vect{t}$ has all co-ordinates bounded by $2^{O(n^2)}$. 
\begin{lemma}[Bounding the coefficients of closest lattice vector]\label{prop:coeff-bound}
For any positive integers $m,n$, there exists an algorithm that given a basis $\mathbf{B}\in \ratn^{m\times n}$ and target vector $\vect{t}\in \ratn^{m}$ of total bitlength $\eta$ as input, outputs a basis $\mathbf{C}\in \ratn^{m\times n}$, target vector $\vect{t}'\in \ratn^m$  such that $\vect{t}'\in \vect{t}+\cL(\basis)$, $\cL(\mathbf{C})=\cL(\mathbf{B})$, the total bitlength of $\mathbf{C}, \vect{t}'$ is at most $\poly(\eta,m,n)$, and for any $\vect{z}\in \intg^{n}$ if \[ \|\mathbf{C}\vect{z}-\vect{t'}\|_2\: =\:\dist_2\left(\cL(\mathbf{C}),\vect{t'}\right) \] 
then \[\|\vect{z}\|_\infty< 2^{n^2}.\]
The algorithm runs in $\poly(\eta,n,m)$ time and requires $\poly(\eta,n,m)$ space. Furthermore, $\mathbf{C}$ is a $\mathsf{LLL}$-reduced basis of basis $\basis$.
\end{lemma}
\begin{proof}
The algorithm does the following. The algorithm runs the \textsf{LLL} algorithm (from \cref{thm:lll}) on basis $\basis$, and gets an \textsf{LLL} reduced basis  $\mathbf{C}=\begin{bmatrix}\vect{c}_1 & \ldots & \vect{c}_n\end{bmatrix}\in \ratn^{m\times n}$  as output. Then the algorithm computes the Gram-Schmidt Orthogonalization(GSO) of basis $\mathbf{C}$ as $\mathbf{C}^*=[\vect{c}^*_1,\ldots,\vect{c}^*_n]$. 
The algorithm then computes 
$ x_i=\frac{\langle \vect{t},\vect{c}_i^*\rangle }{\langle \vect{c}_i^*,\vect{c}_i^*\rangle }\in \ratn$ for all $i \in [n]$. Finally, the algorithm computes 
\[{\vect{t}'}=\vect{t}-\sum_{i=1}^n w_i\vect{c}_i , \text{ where } w_n=\lfloor x_n \rceil, \forall i<n, w_i=\left\lfloor x_i+\sum_{k=i+1}^n w_k\mu_{ki}\right\rceil\;.\]
The algorithm then outputs $\mathbf{C}, \vect{t}'$. 

The algorithm runs in $\poly(\eta,n,m)$ time and all vectors $\vect{c}_i$ for $1 \le i \le n$, and $\vect{t}'$ can be represented in $\poly(\eta,n,m)$ bits. For more details on the computation time of \textsf{LLL} basis, we refer the reader to \cite{Regev04}. We will now prove that $\vect{t}'\in \vect{t}+\cL(\basis)$ and the coefficients of the closest vector in $\cL(\mathbf{C})$ to $\vect{t}'$ are bounded. %
 
As $\mathbf{C}$ is an \textsf{LLL} reduced basis, from \cref{thm:lll} we get the following conditions: 
\begin{equation}\label{eq:lll-condition1}
\forall i\in [n-1], \|\vect{c}^*_i\|^2\leq 2\|\vect{c}^*_{i+1}\|^2
\end{equation}
and
\begin{equation}\label{eq:lll-condition2}
\forall i> j,\; |\mu_{i,j}|\leq \frac{1}{2} \; .\end{equation}
Notice that we can represent the target vector $\vect{t}'$ as  $\vect{t}'=\sum_{i=1}^n x_i^*\vect{c}_i^* +\vect{c}_{n+1}^*$, where  $\vect{c}_{n+1}^*$ lies in a vector space orthogonal to $\vect{c}_1^*,\ldots,\vect{c}_n^*$ in $\real^{m}$. We emphasize here that $\vect{c}_{n+1}^*$ could be $\vect{0}$ if the target vector lies in the linear span of the basis vectors. 
We note that the coefficients $w_i$ are chosen so that, we get that $\forall i\leq n, x_i^*\in (-1/2,1/2]$.
 Also, $\vect{t}-\vect{t}'\in \cL(\mathbf{C})$ as $w_i$'s are integers, and  
  \begin{equation}\label{eq:distance-upper-bound}
\dist(\cL(\mathbf{B}),\vect{t})=\dist(\cL(\mathbf{C}),\vect{t})=\dist(\cL(\mathbf{C}),{\vect{t}'})\leq \|{\vect{t}'}\|
\end{equation}

Let $\vect{v}=\sum\limits_{i=1}^n z_i\vect{c}_i$ be a closest lattice vector to target ${\vect{t}'}$.  We prove by induction on $i$ that $|z_{n-i+1}| \le 2^{n \cdot i}$. Without loss of generality, we assume $\vect{c}_{n+1}^* = \vect{0}$. This does not change the closest vector in the lattice since $\vect{c}_{n+1}^*$ is orthogonal to the lattice. 

We first show that $z_n$ is bounded. To see this, note that 
\begin{align*}
|z_n - x_n^*|\|\vect{c}_n^*\|  \le \|\vect{v} - \vect{t}' \|^2 &\le \left(\frac{1}{4}\sum_{i=1}^n \|\vect{c}^*_j\|^2\right)^{1/2} \\ &\le \frac{1}{2}\|\vect{c}^*_n\|(2^{n-1} + 2^{n-2}+ \cdots + 1)\\ & < 2^{n-1}\|\vect{c}^*_n\| \;. 
\end{align*}
Thus, $|z_n| \le \frac{1}{2} + 2^{n-1} < 2^n$, thereby proving the base case $i = 1$. We now assume that $|z_{n-j+1}| < 2^{n \cdot j}$ for $j \le i$.

For any fixed $z_n, \ldots, z_{n-i+1}$, we now bound $|z_{n-i}|$ corresponding to the vector $\vect{v}$ that minimizes $\|\vect{v} - \vect{t}' \|$. Let $\vect{u} = \sum_{\ell = n-i+1}^n z_\ell \vect{c}_\ell$. For any vector $\vect{x}$, let $\pi(\vect{x})$ denote the projection of $\vect{x}$ in the linear span of $\vect{c}_1^*, \ldots, \vect{c}_{n-i}^*$. Note that the projection of $\vect{v} - \vect{t}'$ in the linear span of $\vect{c}_{n-i+1}^*, \ldots, \vect{c}_{n}^*$ is the same as that of $\vect{u} - \vect{t}'$. Thus, $\|\pi(\vect{v}-\vect{t}')\| \le \|\pi(\vect{u}-\vect{t}')\|$. This implies that
\begin{align*}
\left|\left(z_{n-i} + \left(\sum_{\ell = n-i+1}^n \mu_{n-i,\ell}z_\ell\right) - x_{n-i}\right)\right|^2 \|\vect{c}_{n-i}^*\|^2 & \le  \|\pi(\vect{v}-\vect{t}')\|^2 \\ 
&\le \|\pi(\vect{u}-\vect{t}')\|^2 \\
&\le \sum_{k = 1}^{n-i} \left( \sum_{\ell = n-i+1}^n \mu_{\ell, k} z_\ell - x_k\right)^2 \|\vect{c}_k^*\|^2 \\
&< \sum_{k = 1}^{n-i} \left(\frac{2^{n} + 2^{2n} + \cdots + 2^{i n}}{2} + \frac{1}{2}\right)^2 \|\vect{c}_k^*\|^2 \\
&\le \sum_{k = 1}^{n-i} \left(2^{i n}\right)^2 \|\vect{c}_k^*\|^2 \\
&\le \left(2^{i n}\right)^2 \cdot (1+2+\cdots +2^{n-i-1})\|\vect{c}_{n-i}^*\|^2 \\
&\le \left(2^{i n}\right)^2 \cdot 2^n \cdot  \|\vect{c}_{n-i}^*\|^2 \;,
\end{align*}
using \cref{eq:lll-condition1,eq:lll-condition2}. 

Thus, by triangle inequality,
\begin{align*}
|z_{n-i}| &< \frac{|z_n| + \cdots + |z_{n-i+1}|}{2} + \frac{1}{2} + 2^{in + n/2} \\
&< \frac{2^n + \cdots + 2^{in}}{2} + \frac{1}{2} + 2^{in + n/2} \\
& \le 2^{in}  + 2^{in + n/2} \\
& \le 2^{(i+1) n} \;.
\end{align*}
\end{proof}

In the following theorem, we give an instance compression algorithm for $\CVP_2$. We also show a instance compression with better parameters for $\CVP_2$ in \cref{thm:exact-cvp-euclidean} using a Theorem from \cite{FrankT87}.

\begin{theorem}\label{thm:pcp-approx-gapcvp}
For any positive integers $m,n$, and constant $c_1\in \real^{+}$, given a $\left(1+2^{-n^{c_1}}\right)$-$Gap\CVP(\basis,\vect{t},d)$ instance where $ \basis \in \ratn^{m\times n}$ is a basis of a lattice $\cL$, target $\vect{t}\in \ratn^m$ and $d>0$. The bit-length of the input is at most $\eta$. There exists a $\poly(n,m,\eta)$ time probabilistic algorithm that reduces it to a $(r,q)$-$\CVPip$ instance of size at most $O(n^{c_2}\log^2 (n+m+T))$ for constant $c_2=\max\{c_1+3,5\}$.

 Furthermore, `\textsf{YES}' instance always reduces to `\textsf{YES}' instance and `\textsf{NO}' instance reduces to `\textsf{NO}' instance with at least $1-2^{-n^{3}}$ probability $i.e.$ the reduction does not give false negative.  
\end{theorem}

\begin{proof}
Let $\gamma:=1+2^{-n^{c_1}}$ and $r=2^{n^2}$. We are given a basis $\basis \in \ratn^{m\times n}$, target $\vect{t}\in \ratn^m$ and a distance $d>0$ with a promise that either $\dist(\vect{B},\vect{t})> \gamma d $ or $\dist(\vect{B},\vect{t})\leq d$. From Lemma~\ref{prop:coeff-bound}, we can assume that we are given a $\gamma$-$\GapCVP(\mathbf{C},\tilde{\vect{t}},d)$ instance 
such that $\tilde{\vect{t}}\in \vect{t}+\cL(\basis)$, $\cL(\mathbf{C})=\cL(\basis)$ and for all $\vect{z}\in \intg^n$, if \[\|\mathbf{C}\vect{z}-{\tilde{\vect{t}}}\|=\dist(\mathbf{C},{\tilde{\vect{t}}})\] then  \[\vect{z}\in [-r,r]^{n}.\] 

As $\mathbf{C}\in \ratn^{m\times n}$ and $\tilde{\vect{t}}\in \ratn^{m}$, we can scale the basis vector to make all the coordinates integers and it will not even increase the bit-representations. So, without loss of generality, we assume that $\mathbf{C}\in \intg^{m\times n}$ and $\tilde{\vect{t}}\in \intg^{m}$.
Let $c:=\max\{c_1+1,3\}$. First, we reduce the problem to one where in \textsf{YES} instance the distance of target from lattice is at most $2^{4n^c}$. 
Let $n'$ be an integer such that $2^{n'+1}>d\geq 2^{n'}$.
Let's assume that $n'>4n^c$. Later, we will analyze the case when $n'\leq 4n^c$. 
We remove the $n'-4n^c$ least significant bits of basis vectors $\vect{c}_i$'s and target vector $\tilde{\vect{t}}$. Consider   $\mathbf{C}'=\{\vect{c}_1',\ldots,\vect{c}_n'\}$ and $\vect{t}'$ be the vectors after removing the least significant bits \textit{i.e.}
\[\forall i\in [n], \vect{c}_i'=\left\lfloor\frac{1}{2^{n'-4n^c}} \cdot \vect{c}_{i}\right\rfloor \text{ and }  \vect{t}'= \left\lfloor\frac{1}{2^{n'-4n^c}} \cdot \tilde{\vect{t}}\right\rfloor. \]

We define a new measure of distance from the target, where we only focus on the distance of the target vector from the integer combination of basis vector whose coefficients are less than $2^{n^2}$.  
\[\dist^*({\mathbf{C}'},{\vect{t}'}):=\min_{\vect{z}\in [-r,r]^n }\left\{ \|{\mathbf{C}'}\vect{z}-{\vect{t}'}\| \right\}.\]

We know that closest lattice vector of target vector $\tilde{\vect{t}}$ in lattice $\cL(\mathbf{C})$ is of form $\vect{v}=\sum_{i=1}^n z_i\vect{c}_i$ where $\forall i\leq n,|z_i|< r$. 
Therefore, we get 
\begin{align*}
    \left|\left(2^{n'-4n^c}\cdot \dist^*(\mathbf{C}',\vect{t}')\right)-\dist(\cL(\mathbf{C}),\tilde{\vect{t}})\right|&\leq \max_{\vect{z}\in [-r,r]^n}\left\{\left|\left(2^{r}\cdot \|\mathbf{C}'\vect{z}-\vect{t}'\|\right)-\|\mathbf{C}\vect{z}-\tilde{\vect{t}}\|\right|\right\}\\ &\leq m\cdot n\cdot 2^{n'-4n^c}\cdot 2^{n^{2}}\\& < 2^{n'-2n^c}.
    \end{align*}

From triangle inequality, we get, if $\dist(\cL(\mathbf{C}),\tilde{\vect{t}})\leq d$ then $\dist^*(\mathbf{C}',\vect{t}')\leq \frac{d+2^{n'-2n^c}}{2^{n'-4n^c}}$; otherwise
$\dist^*(\mathbf{C}',\vect{t}')$ is greater than $\frac{\gamma d-2^{n'-2n^c}}{2^{n'-4n^c}}$. Let $d':=\frac{d+2^{n'-2n^c}}{2^{n'-4n^c}}\le 2^{4n^c+1}$ and 
our choice of $c$ implies that
\[
 \frac{\gamma d - 2^{n'-2n^c}}{d + 2^{n'-2n^c}} > 1 \;.
\]
When $n'\leq 4n^c$, we take $\mathbf{C}'=\mathbf{C}$, $\vect{t}'=\tilde{\vect{t}}$ and $d'=d<2^{n'+1}\leq  2^{4n^c+1}$. Hence we get basis $\mathbf{C}'\in \intg^{m\times n}$, target $\vect{t}'\in \intg^m$ and number $d'< 2^{4n^c+1}$ such that $\dist^*(\mathbf{C}',\vect{t}')\leq d'$ if $\dist(\basis,\vect{t})\leq d$, otherwise $\dist^*(\mathbf{C}',\vect{t}')> d'$.

Now, we reduce it to a $\CVP$ instance with explicit bound on coefficients. Let $q$ be a prime chosen uniformly at random from $\left[2^{10n^{c}+\alpha},2^{20n^{c}+\alpha}\right]$ where $\alpha=\log^2(n+m+\eta)$.
Let
\[\forall i\leq n,\;
\vect{h}_{i}:=\vect{c}'_i \mod q \text{ , } \mathbf{H}=\{\vect{h}_1,\cdots,\vect{h}_n\} \text{ and } \vect{t}''=\vect{t}'\mod q \]

We will show that if $\dist^*(\mathbf{C}',\vect{t}')\leq d'$ then there exists a vector $\vect{z}\in [-r,r]^n$ such that $\|\mathbf{H}\vect{z}-\vect{t}''\|^2 \mod q\leq (d')^2$. Otherwise (when $\dist^*(\mathbf{C}',\vect{t}')> d'$ ), for all $\vect{z}\in [-r,r]^n$, $\|\mathbf{H}\vect{z}-\vect{t}''\|^2 \mod q> (d')^2$.

First, let's assume that $\dist^*(\mathbf{C}',\vect{t}')\leq d'$. Let $\vect{z}\in \intg^n$ be a vector such that $\|\mathbf{C}'\vect{z}-\vect{t}'\|=\dist^*(\mathbf{C},\vect{t}')$ and $\|\vect{z}\|_\infty < 2^{n^2}$. From the definition of $\dist^*$ there exist such a vector $\vect{z}$.
Therefore, we get $\left\|\mathbf{H'}\vect{z}-(\vect{t}'' \mod q)\right\|^2\mod q = \|(\mathbf{C}'\vect{z}-\vect{t}')\|^2\mod q \leq (d')^2$.

Now, let's assume that $\dist^*(\mathbf{C}',\vect{t}')>  d'$ \textit{i.e.} for all $\vect{z}\in [-r,r]^n$ , $\|\mathbf{C}'\vect{z}-\vect{t}'\|>d'$. 
From a lower bound on the prime number theorem~\cite{franz} we have that number of primes in range $[2^{10n^c+\alpha}, 2^{20n^c+\alpha}]$ is at least
\[ \frac{2^{20n^c+\alpha}\cdot\log 2}{2\cdot(20n^c+\alpha)}-2^{10n^c+\alpha} \ge 2^{19n^c+\alpha/2} \;.\]
Also, for any fixed $\vect{z}\in [-r,r]^n$ and $w\leq (d')^2$, \[\left|\left\|\sum_{i=1}^n z_i\vect{c}_i'-\vect{t}'\right\|_p^p-w\right| \le m\cdot (n+1)\cdot 2^r\cdot 2^{\delta}+(d')^2 \leq 2^{\poly(n,m,\eta)},\] where 
\[\delta=\max\left\{\log|c_{11}'|,\log|c_{12}'|,\cdots, \log|c_{mn}'|, \log|t'_{1}|,\cdots, \log|t'_m|\right\}\leq \poly(n,m,\eta).\] Hence, there are at most $\poly(n,m,\eta)$ distinct primes  that divide $\left|\|\sum_{i=1}^n z_i\vect{c}_i'-\vect{t}'\|_p^p-w\right|$. Hence,  with probability, at most
\[
\frac{\poly(n,m,\eta)}{2^{19n^c+\alpha/2}} \le 2^{-19n^c}\;,
\]
the prime $q$ is such that
 $\|\mathbf{C}'\vect{z}-\vect{t}'\|_p^p-w= 0\mod q$.
Therefore, by union bound over all $w\leq (d')^2$ and $\vect{z}\in [-r,r]^n$, for uniformly sampled prime $q$ we get   
\begin{equation}\label{eq:soundness-distance-bound}
    \Pr\left[\min_{\|\vect{z}\|_\infty < 2^{n^{2}}}\left\{\|(\mathbf{C}'\vect{z}-\vect{t}')\|^2\mod q\right\}\leq (d')^2 \right] 
\leq 2^{(n^{2}) n}\cdot 2^{4n^c+1} \cdot 2^{-19n^c}  < 2^{-13n^3}.
\end{equation}
It implies that with overwhelming probability, for all $\vect{z}\in [-r,r]^n$, $ $
$\|\mathbf{H}\vect{z}-\vect{t}''\|^2\mod q=\|\mathbf{C}'\vect{z}-\vect{t}'\|^2 \mod q> (d')^2$.

Now, we construct $(r,q)$-$\CVPip$ instance of $(\mathbf{H},\vect{t}'')$ by storing the inner products of basis $H$ and target $\vect{t}''$. We reduce the $\gamma$-$\GapCVP$ instance $(\mathbf{B},\vect{t},d)$ to $(r,q)$-$\CVPip$, where given inner product form of $(\mathbf{H},\vect{t}'')$, integers $d''=(d')^2,r$ and $q$, the goal is to distinguish between a \textsf{YES} instance where there exists a $\vect{z}\in [-r,r]^n$ for which $\|\mathbf{H}\vect{z}-\vect{t}''\|^2\mod q$ is at most $d''$ and a \textsf{NO} instance where for all vector $\vect{z}\in [-r,r]^n$, $\|\mathbf{H}\vect{z}-\vect{t}''\|^2\mod q$ is greater than  $d''$. Notice that, in this reduction we don't get false negative. The instance size is at most $(n+1)^2\cdot \log\left(m\cdot q^2\right)=\mathcal{O}(n^{c+2}\log^2 (n+m+\eta))$ bits because each coordinate of basis $\mathbf{H}$ and target $\vect{t}''$ is less than q. 
It completes the proof.

\end{proof}

\section{Instance compression for all even norms}\label{sec:compress_cvp_even_norm}

In this section, we present an instance compression algorithm for $\CVP_p$ for all even $p$. We show this for any constant $c>0$, $(1+\exp(-n^c))$ approximation of $\CVP_p$ problem with additional promise that distance of the target from lattice is bounded by $\exp(n^c)\cdot \lambda_1^{(p)}$. We will first show that, we can bound the coefficient of the closest lattice vector by using this additional promise.

\begin{lemma}[Bounding the coefficients of $\CVP_p^{\phi}$]\label{lem:coeff-bound-p-norm}
For any $m,n\in \intg^+$, (efficiently computable) $\tau=\tau(m,n)>0$, and $p\in [1,\infty]$, there exists a randomized algorithm that given an instance of $\CVP_p^{\tau\lambda_1}$, basis $\mathbf{B}\in \ratn^{m\times n}$ and target vector $\vect{t}\in \ratn^{m}$ of bitlength $\eta$ as input,  outputs a basis $\mathbf{C}\in \ratn^{m\times n}$ and target vector $\vect{t}'\in \ratn^m$  such that $\cL(\mathbf{C})= \cL(\basis)$, $\vect{t}-\vect{t}'\in \cL(\basis)$ and 
for all vector $\vect{z}\in \intg^n$ which satisfies
\[\|\mathbf{C}\vect{z}-\vect{t}'\|_p= \dist_p(\cL(\mathbf{C}),\vect{t}'),\]
we have \[\|\vect{z}\|_\infty< \tau\cdot m\cdot 2^{3n/2}.\]
The algorithm runs in $\poly(\eta,n,m)$ time and requires $\poly(\eta,n,m)$ space. 
\end{lemma}

\begin{proof}
We are given a basis $\mathbf{B}\in \ratn^{m\times n}$ and target vector $\vect{t}\in \ratn^{m}$ which satisfy 
\begin{equation}\label{eq:bdd-condition}
    \dist_p(\cL({\mathbf{B}}),{{\vect{t}}})\leq \tau\lambda_1^{(p)}\left(\cL({\mathbf{B}})\right)
\end{equation}
The algorithm runs the \textsf{LLL} algorithm (from Theorem~\ref{thm:lll}) on basis $\basis$, and gets an \textsf{LLL} reduced basis $\mathbf{C}=\begin{bmatrix}\vect{c}_1 & \ldots & \vect{c}_n\end{bmatrix}\in \ratn^{m\times n}$ as output. Then the algorithm computes Gram-Schmidt Orthogonalization (GSO) of basis $\mathbf{C}$ as  $\mathbf{C}^*=[\vect{c}^*_1,\ldots,\vect{c}^*_n]$.  

The algorithm then computes 
$ x_i=\frac{\langle {\vect{t}},\vect{c}_i^*\rangle }{\langle \vect{c}_i^*,\vect{c}_i^*\rangle }\in \ratn$ for all $i \in [n]$. Finally, the algorithm computes 
\[    {\vect{t}'}={\vect{t}}-\sum_{i=1}^n w_i\vect{c}_i , \text{ where } w_n=\lfloor x_n \rceil, \forall i<n, w_i=\left\lfloor x_i+\sum_{k=i+1}^n w_k\mu_{ki}\right\rceil\;.
\]
The algorithm then outputs $\mathbf{C}, \vect{t}'$. 
The algorithm runs in $\poly(\eta,n,m)$ time and all vectors $\vect{c}_i$ for $1 \le i \le n$, and $\vect{t}'$ can be represented in $\poly(\eta,n,m)$ bits. For more details on this, we refer the reader to \cite{Regev04}. We will now prove that $\dist_p(\cL(\basis),{\vect{t}})=\dist_p(\cL(\mathbf{C},\vect{t}'))$ and the coefficients of the closest lattice vector with respect to  basis $\mathbf{C}$ to target $\vect{t}'$ are bounded.

As $\mathbf{C}$ is a \textsf{LLL} reduced basis, from \cref{thm:lll} we get the following conditions: 
\begin{equation}\label{eqn:lll-condition3}
\forall i\in [n-1], \|\vect{c}^*_i\|^2\leq 2\|\vect{c}^*_{i+1}\|^2
\end{equation}
and
\begin{equation}\label{eqn:lll-condition4}
\vect{c}_i=\vect{c}_i^*+\sum_{j=1}^{i-1}\mu_{ij} \vect{c}^*_j \text{  and  } \forall i> j, |\mu_{i,j}|\leq \frac{1}{2} .\end{equation}

Notice that we can represent the target vector $\vect{t}'$ as  $\vect{t}'=\sum_{i=1}^n x_i^*\vect{c}_i^* +\vect{c}_{n+1}^*$, where  $\vect{c}_{n+1}^*$ lies in a vector space orthogonal to $\vect{c}_1^*,\cdots,\vect{c}_n^*$ in $\real^{m}$. We emphasize here that $\vect{c}_{n+1}^*$ could be $\vect{0}$ if the target vector lies in the linear span of the basis vectors. 
We note that the coefficients $w_i$ are chosen so that, we get that $\forall i\leq n, x_i^*\in (-1/2,1/2]$.
 Also, $\vect{t}-\vect{t}'\in \cL(\mathbf{C})=\cL(\mathbf{B})$ as $w_i$'s are integers, \textit{i.e.} $\dist_p\left(\cL(\mathbf{C}),\vect{t}'\right)=\dist_p\left(\cL(\basis),{\vect{t}}\right)$.  

Let $\cL'=\cL(\mathbf{C})$. As $\vect{c}_1$ is a non-zero lattice vector, we know that $\lambda_1(\cL')\leq \|\vect{c}_1\|_p$. From Equation~\ref{eq:bdd-condition}, we get that
 \begin{equation}\label{eq:updated-bdd-condition}
     \dist_p\left(\cL',\vect{t}'\right)\leq (1+\tau)\lambda_1^{(p)}\left(\cL'\right)\leq \tau\cdot \|\vect{c}_1\|_p.
 \end{equation}

Let $\vect{v}=\sum_{i=1}^n z_i\vect{c}_i$ be a closest lattice vector to target vector ${\vect{t}'}$ in $\ell_p$ norm. We prove by induction on $i$ that $|z_{n-i+1}| \le \tau\cdot m\cdot i\cdot2^{n/2+i}$. For interchanging between different $\ell_p$ norms, we will be using the fact that for any $\vect{u}\in \real^m$ and $q\geq p \geq 1$, $\|\vect{u}\|_p\leq m^{|1/p-1/q|}\|\vect{u}\|_q$.  We rewrite $\vect{v}=\sum_{i=1}^n y_i\vect{c}_i^*$ and get  
\begin{equation}\label{eq:closest-vector-in-GSO}
    \left\|\vect{v}-{\vect{t}'}\right\|_2^2=\sum_{i=1}^n (y_i-x_i^*)^2\|\vect{c}_i^*\|_2^2+\|\vect{c}_{n+1}^*\|_2^2 \leq m\left(\tau\cdot \|\vect{c}_1\|_p\right)^2 \leq m^2\left(\tau\cdot \|\vect{c}_1\|_2\right)^2,
\end{equation}
where second last inequality follows from \cref{eq:updated-bdd-condition}.
Therefore we get 
\begin{equation}\label{eq:bound-y-i}
    \forall i \in [n], |y_i|\leq \tau m\cdot \frac{\|\vect{c}_1\|_2}{\|\vect{c}_i^*\|_2}+\frac{1}{2}\leq \tau m\cdot 2^{\frac{i-1}{2}}+\frac{1}{2} \leq \tau m\cdot 2^{{n}/{2}},
\end{equation}
using \cref{eqn:lll-condition3}.
By using the fact that $z_n=y_n$ we get $|z_n|< \tau m\cdot 2^{n/2}$, thereby proving the base case $i=1$. We now assume that $|z_{n-j+1}|<\tau m\cdot  2^{n/2+j}$ for $j<i$. 
By using \cref{eqn:lll-condition4}, we get 
\begin{align*}
|z_{n-i+1}|&\leq |y_{n-i+1}|+\sum_{k>n-i+1}|\mu_{k,n-i+1}\cdot z_{k}|\\
&\leq |y_{n-i+1}|+\frac{1}{2}\sum_{k>n-i+1} |z_{k}|\\
&\leq \tau m\cdot2^{n/2}+ \frac{\tau m2^{n/2}}{2}(2+ 2^2+2^3+\ldots+ 2^{i-1})\\
& < \tau m\cdot 2^{n/2+i}.
\end{align*}
Hence, the lemma follows.
\end{proof}

Now, we show an instance compression algorithm for $\CVP_p$ when $p$ is an even integer. It follows the proof technique of \cref{thm:pcp-approx-gapcvp}. For completeness, we also give a proof here. We also show a instance compression with better parameters for $\CVP_p$ in \cref{thm:exact-cvp-p-norm} using a Theorem from \cite{FrankT87}.

\begin{theorem}\label{thm:pcp-approx-gapcvp-even-norms}
For any  $m,n\in \intg^{+}$, $ p\in 2\intg^{+}$ and constant $c_1\in \real^{+}$, given a $\left(1+2^{-n^{c_1}}\right)$-$Gap\CVP_p^{\phi}(\basis,\vect{t},d)$ instance where $ \basis \in \ratn^{m\times n}$ is basis of lattice $\cL$, target $\vect{t}\in \ratn^m$, $d>0$ and $\phi=2^{n^{c_1}}\cdot \lambda_1^{(p)}$. The bit-length of the input is at most $\eta$. There exists a $\poly(n,m,\eta)$ time randomized algorithm that reduces it to a $(r,q)$-$\CVPmvp_p$ instance of size  $O(p\cdot n^{p+c_2}\cdot \log(n+m+\eta))$ for constant $c_2=\max\{c_1+2,3\}$. 

 Furthermore, `\textsf{YES}' instance always reduces to `\textsf{YES}' instance and `\textsf{NO}' instance reduces to `\textsf{NO}' instance with at least $1-2^{-n^{c_2}}$ probability $i.e.$ the reduction does not give false negatives. 
\end{theorem}

\begin{proof}
Let $\gamma:=1+2^{-n^{c_1}}$. We are given a basis $\basis \in \ratn^{m\times n}$, target $\vect{t}\in \ratn^m$ and a distance $d>0$ with a promise that either $\dist_p(\basis,\vect{t})\leq d$ or $\dist_p(\basis,\vect{t})> \gamma d$. We are also given a guarantee that $\dist_p(\basis, \vect{t})\leq 2^{n^{c_1}}\lambda_1^{(p)}(\cL(\basis))=(\gamma-1)^{-1}\lambda_1^{(p)}(\cL(\basis))$.  We first apply the algorithm from Lemma~\ref{lem:coeff-bound-p-norm} with $\tau=\frac{1}{\gamma-1}$, and get basis $\mathbf{C}\in \ratn^{m\times n}$ and target vector $\tilde{\vect{t}}\in \ratn^m$ which satisfies $\cL(\mathbf{C})=\cL(\basis)$, $\tilde{\vect{t}}-\vect{t}\in \cL(\basis)$, 

and for all vector $\vect{z}\in \intg^n$ which satisfies
\[\|\mathbf{C}\vect{z}-\tilde{\vect{t}}\|_p= \dist_p\left(\cL(\mathbf{C}),\tilde{\vect{t}}\right),\]
we have 
\begin{equation}\|\vect{z}\|_\infty< \left(\frac{1}{\gamma-1}\right)\cdot m\cdot 2^{3n/2} < 2^{n^{c_1+1}} .\end{equation}
As $\mathbf{C}\in \ratn^{m\times n}$ and $\tilde{\vect{t}}\in \ratn^{m}$, we can scale the basis vector to make all the coordinates integers and it will not even increase the bit-representations. So, without loss of generality, we assume that $\mathbf{C}\in \intg^{m\times n}$ and $\tilde{\vect{t}}\in \intg^{m}$. Let $c:=\max\{c_1+1,2\}$. We define a new measure of distance from the target, where we only focus on the distance of the target vector from the integer combination of basis vector whose coefficients are less than $2^{n^{c}}$.  
\[\dist_p^*(\mathbf{C},\tilde{\vect{t}}):=\min_{\vect{z}\in \intg^n \text{ and } \|\vect{z}\|_\infty< 2^{n^{c}}}\left\{ \|\mathbf{C}\vect{z}-\tilde{\vect{t}}\| \right\}.\]

Hence, if $\dist_p(\basis,\vect{t})\leq  d$ then $\dist_p^*\left(\mathbf{C},\tilde{\vect{t}}\right)\leq d$, otherwise (when $\dist_p(\basis,\vect{t})> \gamma d$ then)  $\dist_p^*\left(\mathbf{C},\tilde{\vect{t}}\right)>\gamma d$.

We first reduce the problem to one where the distance from the target is at most $2^{4 n^c}$. 
Let $n'$ be an integer such that $2^{n'+1}>d\geq 2^{n'}$.
Let's assume that $n'>4n^c$. Later, we will analyze the case when $n'\leq 4n^c$. We remove the $n'-4n^c$ least significant bits of the entries of basis vectors and target vector. Consider the basis $\mathbf{C}'=\{\vect{c}_1',\ldots,\vect{c}_n'\}$ and $\vect{t}'$ be the lattice and target vector after removing the least significant bits \textit{i.e.}
\[\forall i\in [n], \vect{c}_i'=\left\lfloor\frac{1}{2^{n'-4n^c}} \cdot \vect{c}_i\right\rfloor \text{ and } \vect{t}'=\left\lfloor\frac{1}{2^{n'-4n^c}} \cdot \tilde{\vect{t}}\right\rfloor.\]
Therefore, we get 
\[\left|\left(2^{n'-4n^c}\cdot \dist_p^*(\mathbf{C}',\vect{t}')\right)-\dist_p^*(\cL(\mathbf{C}),\tilde{\vect{t}})\right|\leq m\cdot (n+1)\cdot 2^{n'-4n^c}\cdot 2^{n^{c}} < 2^{n'-2n^c}.\]
Hence, if $\dist_p^*(\mathbf{C},\tilde{\vect{t}})\leq d$ then $\dist_p^*(\mathbf{C}',\vect{t}')\leq \frac{d+2^{n'-2n^c}}{2^{n'-4n^c}}$, and when $\dist_p^*(\mathbf{C},\tilde{\vect{t}})>  \gamma d$ then     $\dist_p^*(\mathbf{C}',\vect{t}')$ is  greater than $\frac{\gamma d-2^{n'-2n^c}}{2^{n'-4n^c}}$.
Let $d':=\frac{d+2^{n'-2n^c}}{2^{n'-4n^c}}<2^{4n^c+2}$ and our choice of $c$ implies that
\[
\frac{\gamma d - 2^{n'-2n^c}}{d + 2^{n'-2n^c}} \geq 1 \;.
\]
When $n'\leq 4n^c$, we take $\mathbf{C}'=\mathbf{C}$, $\vect{t}'=\tilde{\vect{t}}$ and $d'=d<2^{n'+1}\leq  2^{4n^c+1}$. Hence we get basis $\mathbf{C}'\in \intg^{m\times n}$, target $\vect{t}'\in \intg^m$ and number $d'< 2^{4n^c+2}$ such that if $\dist(\basis,\vect{t})\leq d$ then $\dist_p^*(\mathbf{C}',\vect{t}')\leq d'$, and if $\dist_p(\mathbf{B},\vect{t})> d$  then $\dist_p^*(\mathbf{C}',\vect{t}')> d'$.

Now, we reduce to $\CVP$ instance with explicit bounds on basis and target vectors coordinate. Let $q$ be a prime chosen uniformly at random from $\left[2^{10n^{c+1}+\alpha},2^{20n^{c+1}+\alpha}\right]$, where $\alpha=\log^2(n+m+\eta)$.
Let
\[\forall i\leq n,\;
\vect{h}_{i}:=\vect{c}'_i \mod q \text{ , } \mathbf{H}=\{\vect{h}_1,\ldots,\vect{h}_n\} \text{ and } \vect{t}''=\vect{t}'\mod q \]

Let $r:=2^{n^{c}}$. We will show that if $\dist_p^*(\mathbf{C}',\vect{t}')\leq d'$ then there exists a vector $\vect{z}\in [-r,r]^n$ such that $\|\mathbf{H}\vect{z}-\vect{t}''\|_p^p \mod q\leq (d')^p$. Otherwise (when $\dist_p^*(\mathbf{C}',\vect{t}')> d'$ ), for all $\vect{z}\in [-r,r]^n$, $\|\mathbf{H}\vect{z}-\vect{t}''\|_p^p \mod q> (d')^p$.

First, let's assume that $\dist_p^*(\mathbf{C},\vect{t}')\leq d'$. Let $\vect{z}\in \intg^n$ be a vector such that $\|\mathbf{C}'\vect{z}-\vect{t}'\|=\dist_p^*(\mathbf{C},\vect{t}')$ and $\|\vect{z}\|_\infty < r$. From the definition of $\dist_p^*$ there exist such a vector $\vect{z}$.
Hence we get  $\left\|\mathbf{H'}\vect{z}-(\vect{t}'' )\right\|_p^p\mod q = \|(\mathbf{C}'\vect{z}-\vect{t}')\|_p^p\mod q \leq (d')^p$.

Now, let's assume that $\dist_p^*(\mathbf{C},\vect{t}')>  d'$ \textit{i.e.} for all $\vect{z}\in [-r,r]^n$ , $\|\mathbf{C}'\vect{z}-\vect{t}'\|_p>d'$. From a lower bound on the prime number theorem~\cite{franz} we know that number of primes in range $[2^{10n^{c+1}+\alpha}, 2^{20n^{c+1}+\alpha}]$ is at least
\[\frac{2^{20n^{c+1}+\alpha}\cdot \log 2}{2\cdot(20n^{c+1}+\alpha)} - 2^{10n^{c+1}+\alpha} \ge 2^{19n^{c+1}+{\alpha}/{2}} \;.\]
Also, for any fixed $\vect{z}\in [-r,r]^n$ and $w\leq (d')^p$, \[\left|\|\sum_{i=1}^n z_i\vect{c}_i'-\vect{t}'\|_p^p-w\right| \le m\cdot (n+1)\cdot 2^r\cdot 2^{\delta}+(d')^p \leq 2^{\poly(n,m,T)},\]
where \[\delta=\max\left\{\log|c_{11}'|,\log|c_{12}'|,\ldots, \log|c_{mn}'|, \log|t'_{1}|,\ldots, \log|t'_m|\right\}\] is bounded by $\poly(n,m,\eta)$. Hence, there are at most $\poly(n,m,\eta)$ distinct primes  that divide $\left|\|\sum_{i=1}^n z_i\vect{c}_i'-\vect{t}'\|_p^p-w\right|$. Hence,  with probability, at most
\[
\frac{\poly(n,m,\eta)}{2^{19n^{c+1}+\alpha/2}} \le 2^{-19n^{c+1}}\;,
\]
the prime $p$ is such that
 $\|\mathbf{C}'\vect{z}-\vect{t}'\|_p^p-w= 0\mod p$.
 
Therefore, by union bound over all $\vect{z}\in [-r,r]^n$ and $w\leq (d')^p$, we get for uniformly sampled $q$, 
\[    \Pr\left[\min_{\|\vect{z}\|_\infty < 2^{n^{c}}}\left\{\|(\mathbf{C}'\vect{z}-\vect{t}')\|_p^p\mod q\right\}\leq (d')^p \right] 
\leq 2^{(n^{c}) n}\cdot 2^{(4n^c+2)p}  \cdot 2^{-19n^{c+1}}  < 2^{-13n^{c+1}}.
\]
Last inequality uses the assumption that $p=o(n)$.
It implies that with overwhelming probability, for all $\vect{z}\in [-r,r]^n$, $ $
$\|\mathbf{H}\vect{z}-\vect{t}''\|_p^p \mod q=\|\mathbf{C}'\vect{z}-\vect{t}'\|_p^p \mod q> (d')^p$

Now, we construct a $(r,q)$-$\CVPmvp_p$ instance of $(\mathbf{H},\vect{t}'')$ by storing $mvp$ form of basis $\mathbf{H}$ and target $\vect{t}''$. Therefore, we get a randomized reduction from the $\gamma$-$\GapCVP_p^{\phi}$ instance $(\mathbf{B},\vect{t},d)$ to  $(r,q)$-$\CVPmvp_p$, where a \textsf{YES} instance of $\gamma$-$\GapCVP$ always reduces to a \textsf{YES} instance  and a \textsf{NO} instance of $\gamma$-$\GapCVP^{\phi}$ reduces to a \textsf{NO} instance with probability $1-2^{(13n^{c+1})}$. In $(r,q)$-$\CVPmvp_p$, given \textsf{mvp} form of $(\mathbf{H},\vect{t}'')$, with integers $d''=(d')^p,r$ and $q$, the goal is to distinguish between a \textsf{YES} instance where there exists a $\vect{z}\in [-r,r]^n$ for which $\|\mathbf{H}\vect{z}-\vect{t}''\|_p^p \mod q$ is at most $d''$ and a \textsf{NO} instance where for all vector $\vect{z}\in [-r,r]^n$, $\|\mathbf{H}\vect{z}-\vect{t}''\|_p^p \mod q$ is greater than  $d''$. The instance size is at most $(n+1)^p\cdot \log\left(m\cdot q^p\right)=\mathcal{O}\left(p\cdot n^{p+c+1}\log (n+m+\eta)\right)$. 

It completes the proof.

\end{proof}

\subsection{Instance compression for SVP}

We show a instance compression algorithm for $\SVP_p$ for even $p$. For $\SVP$ we don't require any additional promise. For this, we will use a reduction from $\gamma$-$\SVP_p$ to $\gamma$-$\CVP_p$ from \cite{GMSS99}.

\begin{theorem}\label{thm:compress-approx-gapsvp-even-norms}
For any  $m,n\in \intg_{+}$, $ p\in 2\intg_{+}$ and constant $c_1\geq 1$, given a $\left(1+2^{-n^{c_1}}\right)$-$Gap\SVP_p(\basis,d)$ instance where $ \basis \in \ratn^{m\times n}$ is basis of lattice $\cL$, and $d>0$. The bit-length of the input is at most $\eta$. There exists a $\poly(n,m,\eta)$ time randomized algorithm that reduces it to a OR of $n$ $(r,q)$-$\CVPmvp_p$ instances of size $O(p\cdot n^{p+c_2}\cdot \log(n+m+\eta))$ for constant $c_2=\max\{c_1+2,3\}$. 

Furthermore, `\textsf{YES}' instance always reduces to `\textsf{YES}' instance and `\textsf{NO}' instance reduces to `\textsf{NO}' instance with at least $1-\exp(-n)$ probability.
\end{theorem}
\begin{proof}
    From \cref{thm:pcp-approx-gapcvp-even-norms}, we known an instance compression algorithm for $\left(1+2^{-n^{c_1}}\right)$-$Gap\CVP_p^{\phi}(\basis',d)$ where $\phi =2^{n^{c_1}}\lambda_1^{(p)}(\cL(\basis'))$. We will use the reduction from \cite{GMSS99} and show that it reduces to OR of $n$ instances of $\left(1+2^{-n^{c_1}}\right)$-$Gap\CVP_p^{\phi}(\basis',d)$. It is a \textsf{YES} instance of $\gamma$-$Gap\SVP$ if and only if at least one instance of $\gamma$-$Gap\CVP^{\phi}$ is a \textsf{YES} instance. It is enough to just store $n$ compressed $\CVPmvp_p$ instances. 

    Without loss of generality assume that $\basis$ is a \textsf{LLL} reduced basis. Let $\basis_i=[\vect{b}_1,\ldots, 2\vect{b}_i,\ldots \vect{b}_n]$ and $\vect{t}_i=\vect{b}_i$. We will show that $(\basis_i,\vect{t}_i,d)$ is a valid instance of $\GapCVP_p^{\phi}$ where $\phi\geq m\sqrt{n}\cdot 2^{3n/4} \lambda_1^{(p)}$. Notice that for all $i\in [n]$ and $\vect{z}\in \intg^n$, $\vect{B}_i\vect{z}-\vect{t}_i\in \cL\setminus \{\vect{0}\}$ and $\lambda_1^{(p)}(\cL(\basis_i))\geq \lambda_1^{(p)}(\cL(\basis))$. Therefore, it is enough to show that $\dist_p(\cL(\basis_i),\vect{t}_i)\leq \|\vect{b}_i\|_p \leq 2^{n^{c_1}}\lambda_1(\cL)$. From \cref{thm:lll}, we get that $\|\vect{b}_1\|_2\leq 2^{n/2}\lambda_1^{2}(\cL)$, $\vect{b}_i=\sum_{j=1}^i \mu_{ij} \vect{b}_j^*$,and  $\forall n\geq i>j\geq 1, \|\mu_{ij}\|\leq 1/2$ and $\|\vect{b}_i^*\|_2^2 \geq 1/2\cdot \|\vect{b}_{i-1}^*\|_2^2$. Therefore, we get 
    \begin{align*}
        \|\vect{b}_i\|_p^2\leq m\cdot \|\vect{b}_i\|_2^2\leq m\left(\sum_{j=1}^i\|\mu_{ij}\|^2\|\vect{b}_j^*\|_2^2\right) \leq m\cdot\left(\sum_{j=1}^{i}\|\vect{b}_j^*\|_2^2\right)&\leq m\cdot i\cdot 2^{i/2}\cdot \|\vect{b}_1\|_2^2 \\
        &\leq  mn\cdot 2^{3n/2} \left(\lambda_1^{(2)}(\cL)\right)^2\\
        &\leq m^2n\cdot 2^{3n/2} \left(\lambda_1^{(p)}(\cL)\right)^2.
    \end{align*}
    Hence $(\basis_i,\vect{t}_i,d)$ is a valid instance of $\GapCVP_p^{\phi}$ where $\phi\geq m\sqrt{n}\cdot 2^{3n/4} \lambda_1^{(p)}$.

    Now we show that if $\lambda_1(\cL)\leq d$ then there exist an $i\in [n]$ for which $\dist_p(\cL(\basis_i),\vect{t}_i)\leq d$. Let $\vect{v}=\sum \vect{z}_i\vect{b}_i$ be a shortest non-zero lattice vector in $\ell_p$ norm. We use the fact the fact that there exists an index $j$ such that $z_j$ is odd integer. Otherwise $\vect{v}/2$ is also a lattice vector which contradicts the assumption that $\vect{v}$ is shortest lattice vector. Therefore, we get that $\dist_p(\cL(\basis_j),\vect{t}_j)=\lambda_1(\cL) \leq d$. For other direction, if $\lambda_1(\cL)>\gamma d$ then for all $i$, $\dist_p(\cL(\basis_i),\vect{t}_i)>\gamma d$. It directly follows from the fact that for all $\vect{x}\in \intg^n$, $\basis_i\vect{x}-\vect{t}_i\in \cL\setminus \{\vect{0}\}$.
    We use \cref{thm:pcp-approx-gapcvp-even-norms} and store $n$ instances of a variant of $\CVPmvp_p$. It completes the proof.   
\end{proof}
\section{Compression for exact-CVP in Even norms}\label{sec:compresssion_exact}

We will use the following theorem from \cite{FrankT87}.

\begin{theorem}\cite{FrankT87}[Theorem~3]\label{thm:det_compress_FT87}
    For any positive integers $m,N$, given set of integers $\Sigma=\{a_1,\ldots, a_m\}$ there exists a polynomial time algorithm that outputs $\Sigma'=\{a'_1,\ldots, a'_m\}$ where $\forall i\in[m]$,$|a'_i|\leq 2^{4m^3}N^{m(m+2)}$ and satisfy the following: for all $\vec{z}\in \intg^{m}$, if $\|\vec{z}\|_{1}\leq N-1$ then \[{\rm sign}\left(\sum_{i=1}^m z_ia_i\right)={\rm sign}\left(\sum_{i=1}^m z_ia'_i\right).\]
\end{theorem}

Now, we will show a deterministic compression for exact-CVP.

\begin{theorem}\label{thm:exact-cvp-euclidean}
    For any positive integers $m,n$, given a $\CVP(\basis,\vec{t},d)$ instance of bitlength $\eta$ where $\basis \in \intg^{m\times n}$ is a basis of a lattice $\cL$, target $\vec{t}\in \intg^m$ and $d>0$. There exists a $\poly(n,m,\eta)$ time algorithm that reduces it to a $2^{n^2}$-$\CVPip$ instance of size $O(n^8)$ bits.
\end{theorem}

\begin{proof}
   We are given a basis $\basis \in \intg^{m\times n}$, target $\vec{t}\in \intg^m$ and a distance $d>0$. Without loss of generality, by applying \cref{prop:coeff-bound}, we can assume that  for any $\vec{z}\in \intg^n$,
   \begin{equation}\label{label:bound_on_coeff}
   \|\basis\vec{z}-\vec{t}\|=\dist(\basis,\vec{t}) \implies \vec{z}\in \left[-2^{n^2},2^{n^2}\right]^n. 
  \end{equation} 
   
     Let $\forall i,j\in [n]$, $\alpha_{i,j}:=\langle \vec{b}_i,\vec{b}_j\rangle $ and $\beta_{i}:= -\langle \vec{b}_i,\vec{t}\rangle$, $\beta_{n+1}:=\langle\vec{t},\vec{t}\rangle$. It is easy to see that for all $\vec{z}\in \intg^n$,
   \[\|\basis \vec{z}-\vec{t}\|_2^2=\sum_{i=1}^n\left(\sum_{j=1}^n (z_i\cdot z_j)\alpha_{i,j}+z_i \beta_{i}\right)+\beta_{n+1}.\]
   Notice that this expression is linear in variables $\{z_1z_1,z_1z_2,\ldots,z_nz_n,z_1,\ldots,z_n\}\in \intg^{n^2+n}$ and from \cref{label:bound_on_coeff} we are only interested in the evaluation of above expression when each component of this vector has absolute value at most $2^{2n^2}$.
   
   Now we will apply the deterministic compression algorithm from \cite{FrankT87}. Let \[\Sigma=\{ \alpha_{i,j},\beta_{i},\beta_{n+1}, d| \forall i,j\in [n]\} \in \intg^{n^2+n+2} \text{ and } N=(n^2+n+2)2^{2n^2}+1\;.\] By using \cref{thm:det_compress_FT87}, we get $\Sigma'=\{ \alpha'_{i,j},\beta'_{i},\beta'_{n+1}, d'| \forall i,j\in [n],\}\in \left[ 2^{7n^6}\right]^{n^2+n+2}$  which satisfies, for any $\vec{z}\in [-2^{n^2},2^{n^2}]^n$, 
   \[\sum_{i=1}^n\left(\sum_{j=1}^n (z_i\cdot z_j)\alpha_{i,j}+z_i \beta_{i}\right)+\beta_{n+1}\leq d\]
   if and only if 
   \[\sum_{i=1}^n\left(\sum_{j=1}^n (z_i\cdot z_j)\alpha'_{i,j}+z_i \beta'_{i}\right)+\beta'_{n+1}\leq d'.\]
    Hence it reduces to $2^{n^2}$-$\CVPip$ instance with input as inner products $\forall i,j\in [n]$, $\alpha'_{i,j},\beta'_{i},\beta_{n+1}$ and number $d'$. It is also easy to compute that instance size is $O(n^8)$ bits.    
   
\end{proof}

We also generalize this result to compression of $\CVP_p$ instance for even integer $p$. It requires an additional condition that the distance from the lattice is bounded by some factor times the length of shortest non-zero lattice vector.

\begin{theorem}\label{thm:exact-cvp-p-norm}
For any  $m,n\in \intg^{+}$, $ p\in 2\intg^{+}$ and constant $c_1\in \real^{+}$, given a $\CVP_p^{\phi}(\basis,\vect{t},d)$ instance where $ \basis \in \ratn^{m\times n}$ is basis of lattice $\cL$, target $\vect{t}\in \ratn^m$, $d>0$ and $\phi=2^{n^{c_1}}\cdot \lambda_1^{(p)}$. The bit-length of the input is at most $\eta$. There exists a $\poly(n,m,\eta)$ time algorithm that reduces it to a $\tau$-$\CVPmvp_p$ instance of size  $O(n^{4p}+n^{3p}(n^{c_1}+\log m))$ where $\tau=2^{n^{c_1}}\cdot m\cdot 2^{3n/2}$.
\end{theorem}

\begin{proof}
   We are given a basis $\basis \in \intg^{m\times n}$, target $\vec{t}\in \intg^m$ and a distance $d>0$. Without loss of generality, by applying \cref{lem:coeff-bound-p-norm}, we can assume that  for any $\vec{z}\in \intg^n$,
   \begin{equation}\label{label:bound_on_coeff_p_norm}
   \|\basis\vec{z}-\vec{t}\|=\dist(\basis,\vec{t}) \implies \vec{z}\in \left[-\tau,\tau\right]^n,
  \end{equation} 
   where $\tau= 2^{n^c_1}\cdot m\cdot 2^{3n/2}$.
    Let $\vec{b}_{n+1}=\vec{t}$, $\forall (i_1,\ldots,i_p)\in [n+1]^p$
\[\mvp(i_1,\ldots,i_p)= \sum_{k=1}^m \left(\prod_{j=1}^p b_{kj}\right).\]
   
From \cref{lem:cvp-mvp}, we get that for all $\vec{z}\in \intg^n$,
   \[\|\basis \vec{z}-\vec{t}\|_p^p=\sum_{(j_1,\ldots,j_p)\in [n+1]^p} (z_{j_1}\cdots z_{j_p})\mvp(j_1,\ldots,j_p),\]
   where $z_{n+1}=-1$.
   Notice that this expression is linear in variables $\{\forall (j_1,\ldots,j_p)\in [n+1]^p, z_{j_1}\ldots z_{j_p}\}\in \intg^{(n+1)^p}$ and from \cref{label:bound_on_coeff_p_norm} we are only interested in the evaluation of above expression when each coordinate of coefficient vector has absolute value at most $\tau^{p}$ i.e. $(z_{j_1}\cdots z_{j_p})\leq\tau^p$.
   
   Now we will apply the deterministic compression algorithm from \cite{FrankT87}. Let \[\Sigma=\{ \mvp(j_1,\ldots,j_p), d| \forall (j_1,\ldots,j_p)\in [n+1]^p\} \in \intg^{(n+1)^p+1} \text{ and } N=((n+1)^p+1)\cdot \tau^{p}+1\;.\] Let
  $M=2^{4((n+1)^p+1)^3}\cdot N^{((n+1)^p+1)((n+1)^p+3)}$. By using \cref{thm:det_compress_FT87}, we get \[\Sigma'=\{ \alpha_{i_1,\ldots,i_p}, d'| \forall (i_1,\ldots,i_p)\in [n+1]^p\}\in \left[ M\right]^{(n+1)^p+1}\] which satisfies, for any $\vec{z}\in [-\tau,\tau]^n$, $z_{n+1}=-1$, 
   \[\sum_{(j_1,\ldots,j_p)\in [n+1]^p} (z_{j_1}\cdots z_{j_p})\mvp(j_1,\ldots,j_p)\leq d\]
   if and only if 
   \[\sum_{(j_1,\ldots,j_p)\in [n+1]^p} (z_{j_1}\cdots z_{j_p})\alpha_{j_1,\ldots,j_p}\leq d'.\]
    Hence it reduces to $\tau$-$\CVPmvp_p$ instance with input as multi-vector products $\forall (j_1,\ldots,j_p)\in [n+1]^p$, $\alpha_{j_1,\ldots,j_p}$ and number $d'$. It is also easy to compute that instance size is $(\log M)\cdot ((n+1)^p+1)=O(n^{4p}+n^{3p}(n^{c_1}+\log m))$ bits.    
   
\end{proof}

\begin{corollary}\label{cor:exact-svp}
    For any  $m,n\in \intg_{+}$, $ p\in 2\intg_{+}$ and constant $c_1\geq 1$, given a $\SVP_p(\basis,d)$ instance where $ \basis \in \ratn^{m\times n}$ is basis of lattice $\cL$, and $d>0$. The bit-length of the input is at most $\eta$. There exists a $\poly(n,m,\eta)$ time  algorithm that reduces it to a OR of $n$ $\tau$-$\CVPmvp_p$ instances of size $O(n^{4p}+n^{3p}(n+\log m)$ where $\tau=m^2\sqrt{n}2^{9n/4}$. 
\end{corollary}

Proof exactly follows the arguments of \cref{thm:compress-approx-gapsvp-even-norms} and uses \cref{thm:exact-cvp-p-norm}.

\section{Implication to SETH hardness of CVP}
\label{sec:barriers}

In this section, we show that it is not possible to get a polynomial time Turing reduction from $k$-$\SAT$ to $\CVP_2$ instance of $n^c$ rank lattice unless polynomial hierarchy collapses to third level. We also generalize this result for $\CVP_p$ (distance guarantee)  for even $p$.  We extend this result for exponential time reduction with the restriction that reduction will only make fixed polynomial number of calls to $\CVP_p$ oracle. We also give barriers for randomized polynomial time reductions. However, we are only able to show it for non-adaptive reductions.

\begin{theorem}\label{thm:SAT-to-CVP2}
For any constants $c, c_1>0 $,  there exists a constant $k_0$ such that for any $k> k_0$ there does not exists a polynomial time probabilistic reduction with no false negatives from $k$-$\SAT$ on $n$-variables to $\CVP_2$ on $n^{c_1}$ rank lattice that makes at most $n^{c}$ calls to $\CVP_2$ oracle, unless $\coNP \subseteq \NPPoly$.
\end{theorem}
\begin{proof}
    Given a $\CVP_2$ instance of $n^{c_1}$ rank lattice, by \cref{thm:exact-cvp-euclidean}, we get a compressed instance of $\CVPip$ of size $\tilde{\mathcal{O}}(n^{8c_1})$. Let $k_0= c+8c_1$. Let's assume that for some $k>k_0$ there exist a polynomial time probabilistic reduction without false negative from $k$-$\SAT$ on $n$-variables to $\CVP_2$ on $n^{c_1}$ rank lattice and it makes at most $n^{c}$ calls to $\CVP_2$ oracle. The reduction can also be seen as an oracle communication protocol for $k$-$\SAT$ where for each call to $\CVP_2$ instance, first player send the compressed instance to second play (which is computationally unbounded). From definition of oracle communication protocol, the cost of this protocol is at most $n^c\cdot \tilde{\mathcal{O}}(n^{8c_1})$. Therefore, by \cref{thm:compress-SAT}, we get $\coNP \subset \NPPoly$ as $k>k_0=c+8c_1$.
    
    Notice that, the above arguments holds for \emph{adaptive} reduction for $k$-$\SAT$ to $\CVP_2$. This completes the proof.
\end{proof}

We also get the similar result for $\CVP_p$ for all even positive integer $p$.
\begin{theorem}\label{thm:SAT-to-CVPp}
For any constants $c, c_1,c_2>0$ and $p\in 2\intg^{+}$,  there exists a constant $k_0$ such that for any $k>k_0$ there does not exists a polynomial time probabilistic reduction with no false negatives from $k$-$\SAT$ on $n$-variables to $\CVP^{\phi}_p$ on $n^{c_2}$ rank lattice where $\phi=2^{n^{c_1}}\cdot\lambda_1^{(p)}$, that makes at most $n^{c}$ calls to $CVP^{\phi}_p$ oracle, unless $\coNP \subseteq \NPPoly$.
\end{theorem}

\begin{proof}
  Given a $\CVP_p^{\phi}$ instance of $n^{c_2}$ rank lattice, by \cref{thm:exact-cvp-p-norm}, we get a compressed instance of $\CVPmvp$ of size $\tilde{\mathcal{O}}(n^{c_3})$ where $c_3=c_2\cdot \max\{4p,\;3p+c_1\}$. Let $k_0= c+c_3$. Let's assume there exist a polynomial time probabilistic reduction without false negative from $k$-$\SAT$ on $n$-variables to $\CVP^{\phi}_p$ on $n^{c_2}$ rank lattice and it makes at most $n^c$ calls to $\CVP^{\phi}_p$ oracle. The reduction can also be seen as an oracle communication protocol for $k$-$\SAT$ where  for each call to $\CVP^{\phi}_p$ instance, first player send the compressed instance to second play (which is computationally unbounded). From definition of oracle communication protocol, the cost of this protocol is at most $n^c\cdot \tilde{\mathcal{O}}(n^{c_3})$. Therefore, by \cref{thm:compress-SAT}, we get $\coNP \subset \NPPoly$ as $k>k_0= c+c_3 $.
    
Notice that, the above arguments also holds for adaptive reduction for $k$-$\SAT$ to $\CVP^{\phi}_p$. This completes the proof.
\end{proof}

For $\SVP_p$, we get the following result.
\begin{theorem}\label{thm:SAT-to-SVPp}
For any constants $c, c_1>0$ and $p\in 2\intg^{+}$,  there exists a constant $k_0$ such that for any $k>k_0$ there does not exists a polynomial time probabilistic reduction with no false negatives from $k$-$\SAT$ on $n$-variables to $\SVP_p$ on $n^{c_1}$ rank lattice that makes at most $n^{c}$ calls to $\SVP_p$ oracle, unless $\coNP \subseteq \NPPoly$.
\end{theorem}

Proof directly follows from \cref{cor:exact-svp} and \cref{thm:compress-SAT}; it follows the same arguments as above.

We extend these barrier for exponential time reduction. Specifically, we get the following results.

\begin{theorem}\label{cor:exptime-SAT-to-CVP}
For any constants $c, c_1>0 $,  there exists a constant $k_0$ such that for any $k>k_0$ and $T>0$, there does not exists a probabilistic reduction without false negative from $k$-$\SAT$ on $n$-variables to $\CVP_2$ on $n^{c_1}$ rank lattice, in time $T$ and reduction makes at most $n^{c}$ calls to $\CVP_2$ oracle,  unless $\coNP \subseteq \frac{\NTIME(T\cdot \poly)}{\Poly}$.
\end{theorem}

Proof follows directly from \cref{thm:randomized-oneside-compress} and \cref{thm:exact-cvp-euclidean}; it follows the same arguments as \cref{thm:SAT-to-CVP2}. 

By \cref{thm:exact-cvp-p-norm}, we also get following result for $\ell_p$ norms for all positive even integer $p$.

\begin{theorem}\label{cor:exptime-SAT-to-CVPp}
For any constants $c, c_1,c_2>0 $ and $p\in 2\intg^{+}$,  there exists a constant $k_0$ such that for any $k>k_0$ and $T>0$, there does not exists a probabilistic reduction with no false negatives from $k$-$\SAT$ on $n$-variables to $\CVP_p^{\phi}$ on $n^{c_2}$ rank lattice where $\phi=2^{n^{c_1}}\cdot \lambda_1^{(p)}$, in time $T$ and reduction makes at most $n^{c}$ calls to $\CVP_p^{\phi}$ oracle,  unless $\coNP \subseteq \frac{\NTIME(T\cdot \poly)}{\Poly}$.
\end{theorem}

We also give a barrier for randomized polynomial time reduction with both side error from $k$-$\SAT$ to $\CVP$ in even norm. We are only able  to show this barrier for \emph{non-adaptive} reductions. We require the following result from \cite{drucker2015new}.

\begin{theorem}\cite{drucker2015new}[Theorem~7.1]\label{thm:randomized-compress-SZK}
    Let $L$ be any language and $t_1=t_1(n),t_2=t-2(n)>0$. Suppose that there exists a probabilistic polynomial time instance compression of  OR($L$) such that it reduces problem of OR of $t_1$ instances of $L$ of length $n$ to instance of some language $L'$ of bitlength $t_2$  with error bound $\eps(n) < 1/2$. Let
\[\delta=\min \left\{ \sqrt{\frac{\ln 2}{2}\cdot \frac{t_2+1}{t_1}}, 1-2^{-\frac{t_2}{t_1}-3}  \right\}\]
If for some constant $c>0$, we have 
\[ (1-2\eps(n))^2-\delta \geq \frac{1}{n^c}\]
then there is a Karp reduction from $L$ to a problem in \textsf{SZK}. The reduction is computable in non-uniform polynomial time; in particular this implies $L\in \NPPoly \cap \coNP/\mathsf{Poly}$ 
\end{theorem}

\begin{theorem}\label{thm:rand-poly-time-compress-CVP2}    
For any constants $c, c_1>0 $,  there exists a constant $k_0$ such that for any $k>k_0$ there does not exists a (non-adaptive) randomized polynomial time reduction from $k$-$\SAT$ on $n$-variables to $\CVP_2$ on $n^{c_1}$ rank lattice with constant error bound 
 and reduction make at most $n^c$ calls to $\CVP_2$ oracle, unless  there are non-uniform, statistical zero-knowledge proofs for all languages in $\NP$.
\end{theorem}
\begin{proof}
Let $c_2:=8c_1$ and $k_0= c+c_2$. Let's assume that there exists a non-adaptive randomized polynomial time reduction from $k$-$\SAT$ on $n$-variables to $\CVP_2$ on $n^{c_1}$ rank lattice with constant error bound and the reduction make at most $n^c$ calls to $\CVP_2$ oracle. Using \cref{thm:exact-cvp-euclidean}, for $\CVP_2$ on $n^{c_1}$ rank lattice, we get instance compression of size $\tilde{\mathcal{O}}(n^{c_2})$. As the reduction from $k$-$\SAT$ to $\CVP_2$ is non-adaptive, in polynomial time we can identify the $\CVP_2$ instances to which reduction will make oracle calls. Therefore, by storing corresponding compressed $\CVP_2$ instances, we get probabilistic instance compression of $k$-$\SAT$ of size $\tilde{O}(n^{c+c_2})= O(n^{k_0})$.

Now, from Lemma~\ref{lem:VCover-SAT}, we get an instance compression for $k$-Vertex cover of $\mathcal{O}(n^{k_0})$ size. Notice that, by a trivial reduction between $k$-Vertex cover and $k$-Clique, we also get $\mathcal{O}(n^{k_0})$ bit-size instance compression for $k$-Clique. Therefore, by Lemma~\ref{lem:OR-SAT-Clique}, there is an instance compression for OR($3$-$\sat$) of bitlength $\mathcal{O}\left(\left(n\cdot \max\{n,t^{1/k+o(1)}\}\right)^{k_0}\right)$. As $k_0$ is a constant less than $k$, if we take $t$ as sufficiently large polynomial in $n$ we get a $\mathcal{O}(t\log t)$ bitlength instance compression of OR($3$-$\sat$) with constant error bound. Notice that this sequence of reductions are deterministic and preserves the error bound. Therefore, by \cref{thm:randomized-compress-SZK}, we get a non-uniform polynomial time Karp reduction from $3$-$\SAT$ to a problem in $\mathsf{SZK}$. Hence, it implies that there are non-uniform, statistical zero-knowledge proofs for all languages in $\NP$. It completes the proof.
   
\end{proof}

By \cref{thm:exact-cvp-p-norm} and following same arguments as above we get the following result for $\CVP_p$ for all even positive integers $p$.

\begin{theorem}\label{cor:rand-poly-time-compress-CVPp}
For any constants $c, c_1,c_2>0 $,  there exists a constant $k_0$ such that for any $k>k_0$ there does not exists a (non-adaptive) randomized polynomial time reduction from $k$-$\SAT$ on $n$-variables to $\CVP_p^{\phi}$ on $n^{c_2}$ rank lattice with constant error bound where $\phi=2^{n^{c_1}}\lambda_1^{(p)}$,and reduction make at most $n^c$ calls to $\CVP_p^{\phi}$ oracle, unless there are non-uniform, statistical zero-knowledge proofs for all languages in $\NP$.
\end{theorem}
    
\begin{remark}
 Note that, in this paper, all the barriers for reduction from $k$-$\SAT$ to $\CVP_p^{\phi}$ also holds for reduction from $k$-$\SAT$ to $\mathsf{SVP}_p$ because of an efficient reduction from $\mathsf{SVP}_p$ to $\CVP_p^{\phi}$~\cite{GMSS99}.
 \end{remark}

\section{Barrier for \textsf{SETH}-hardness of \textsf{Subset-Sum}}\label{sec:subsetsum}

\textsf{Subset-Sum} is one of the most extensively studied problem in computer science. Showing a fine-grained hardness of \textsf{Subset-Sum} based on $\SETH$ is an important open problem. Harnik and Naor~\cite{HN10}, gave an algorithm for instance compression of arbitrary \textsf{Subset-Sum} instance. In this section, we will describe the consequences of this compression to get a reduction from  $k$-$\SAT$ to \textsf{Subset-Sum}.

\begin{lemma}\label{lem:subsetsum-compress}\cite{HN10}[Claim~2.7]\label{lem:compress-subset-sum}
For any positive integers $n,m$, there exists a randomized polynomial time algorithm that compresses any arbitrary \textsf{Subset-Sum} instance $(a_1, \ldots, a_n, t)$ on $n$ numbers with \[\eta := 
\left\lceil \max\left\{\log_2 |a_1|, \ldots, \log_2 |a_n|,\log_2 |t|\right\} \right\rceil \;, \] to $O(n^2+n\log \eta)$ bits. 
Furthermore, the reduction does not give false negative.
\end{lemma}

\begin{theorem}\label{thm:barrier-for-subset-sum}
For any constants $c,c'>0$,there exists a constant $k_0$ such that for any $k>k_0$ there does not exists a polynomial time probabilistic reduction with no false negatives from $k$-$\sat$ on $n$-variable to \textsf{Subset-SUM} on $n^{c'}$ numbers which makes at most $O(n^c)$ calls to \textsf{Subset-Sum} oracle, unless $\coNP \subseteq \NPPoly$.
\end{theorem}

\begin{proof}
  Given a \textsf{Subset-Sum} instance of $n^{c'}$ numbers, by \cref{lem:subsetsum-compress}, we can get a compressed instance of size ${\mathcal{O}}(n^{2c'})$. Let $k_0=c+2c'$. Let's assume there exist a polynomial time probabilistic reduction without false negative from $k$-$\SAT$ on $n$-variables to \textsf{Subset-Sum} instance on $n^{c'}$ numbers and the reduction only makes $O(n^c)$ calls to \textsf{Subset-Sum} oracle. Notice that, the reduction can also be seen as an oracle communication protocol where  for each call to \textsf{Subset-Sum} instance, first player send the compressed instance to second play (which is computationally unbounded). From definition of oracle communication protocol, the cost of this protocol is at most $n^c\cdot \mathcal{O}(n^{2c'})=\mathcal{O}(n^{k_0})$. Therefore, by \cref{thm:compress-SAT}, we get $\coNP \subset \NPPoly$ as $k>c+c' $.
  
Notice that, the above arguments holds for \emph{adaptive} reductions from $k$-$\SAT$ to \textsf{Subset-Sum}. This completes the proof.
\end{proof}

\begin{remark}
    Note that, similar to \cref{cor:exptime-SAT-to-CVP,thm:rand-poly-time-compress-CVP2}, we can also get the same barriers for reductions from $k$-$\sat$ to \textsf{Subset-Sum}. 
\end{remark}

\bibliographystyle{alpha}

\section*{Acknowledgments} 
The authors thank Huck Bennett, Zvika Brakerski, Alexander Golovnev, Zeyong Li, Noah Stephens-Davidowitz, and Prashant Nalini Vasudevan for helpful discussions. In particular, we thank Alexander Golovnev for many helpful comments on an earlier draft of this work. We also thank FOCS reviewers for pointing out the instance compression algorithm for exact-CVP. Divesh Aggarwal was supported by the bridging grant at Centre for Quantum Technologies titled ``Quantum algorithms, complexity, and communication". Rajendra Kumar was supported by the European Union Horizon 2020 Research and Innovation Program via ERC Project REACT (Grant 756482).
\bibliography{ref}
\appendix
\section{Proof of Theorem~\ref{thm:randomized-oneside-compress}}\label{sec:compression-impossibility}

\begin{lemma}\label{lem:compression-implication}
Let $L$ be a language and $t:\nat \rightarrow \nat\setminus \{0\}$ be polynomially bounded in $n$ such that the problem of deciding whether at least one of $t$ inputs of length at most $n$ belongs to $L$ has an oracle communication protocol, of cost $\mathcal{O}(t\log t)$, where the first player runs in time $T$. Then $\overline{L}\in \cc{NTIME(\poly(n)\cdot T)}/\poly$. 
\end{lemma}
\begin{proof}
Let $P$ be the oracle communication protocol for language $L$ that runs in time $\poly(n)\cdot T$ such that the output is a deterministic function of the communication transcript. For input $x$, protocol $P$ makes queries to the second party (which we sometimes call the oracle); and receives some outputs from the second party. We use $C$ to denote the communication transcript between the two parties. The cost of the protocol is the number of bits of communication from the first player to the second player. First player is allowed to use randomness and  protocol always output accept (outputs \textsf{YES}) if instance belongs to the language $L$.

We will use the following equivalence: an input $x$ is in $\overline{L}$ if and only if there exist a sequence $(x_2,\ldots, x_t)$ and randomness $r$ such that $P(x,x_2,\ldots, x_t;r)$ rejects (outputs \textsf{NO}). It follows from the fact that protocol does not give false negative. Let $s=\poly(n,t)$ be the number of random bits required to execute the oracle communication protocol. We will show existence of a polynomial size advice string $A_n$ which contains a subset of the transcripts of the protocol $P$ such that for every $x\in \overline{L}$ there exists $x_2,\ldots, x_t$ and randomness $r$ such that the communication transcript $C$ of protocol $P$ on input $(x,x_2,\ldots, x_t;r)$ is in $A_n$ and $P(x,x_2,\ldots, x_t;r)$ rejects. Before showing the existence of the advice string, we show why this is sufficient. By using such an advice string $A_n$, we get the following algorithm for input $x$:
\begin{enumerate}
    \item Guess a sequence $(r,x_2,\ldots,x_t)$, where each $x_i$ is of length $n$ and randomness $r$ is of length $s$.
    
    \item Check whether the communication transcript $C$ of protocol $P$ on input $(x,x_2,\ldots,x_t; r)$ is consistent with a transcript in $A_n$ and $P(x,x_2,\ldots,x_t;r)$ rejects. If so then accept, otherwise reject.
\end{enumerate}

To check the consistency of protocol $P$ on input $(x,x_2,\ldots,x_t; r)$ with any transcript $\tau$, we check the input of the first player to the second player is consistent with transcript $\tau$ and whenever the first player expects an output from the second player we give the desired output by using the transcript $\tau$. The correctness of the protocol follows from the equivalence as mentioned earlier. 

In the rest of the proof, we will show that there exists a polynomial-size advice string $A_n$.

We find the set $A_n$ in the greedy way. Notice that any transcript $\tau$ of the protocol $P$ on input $(x,x_2,\cdots, x_t;r)$ is a deterministic function of the bits sent by the first player to second player. Let us assume that the cost of oracle communication protocol is $c=\mathcal{O}(t\log t)$. It implies that there are at most $2^{c}$ many distinct rejecting transcripts. 

For any rejecting transcript $\tau$, and any $x \in \overline{L}$, we say that $\tau$ covers $x$ if there exists $x_2\ldots, x_t$ and $r$ where $x_i\in \overline{L}$ and $r_i\in \{0,1\}^{s}$ such that $\tau$ is a  transcript for $P(x, x_2, \ldots, x_t; r)$. We will iteratively add rejecting transcripts to $A_n$ such that they eventually cover all $x\in \overline{L}$.

Let $F$ be the set of all $x\in \overline{L}$ that have so far not been covered.  We known that for every $t$-tuple of instances in $F^t$ for some randomness $r$, there exists a rejecting transcript (in fact, there exist many rejecting transcripts). So, since there are at most $2^c$ transcripts in total, there exists a transcript $\tau$ that is a rejecting transcript for at least $|F|^t/2^c$ tuples of instances. Let $G$ be the subset of $F$ that is covered by $\tau$. Thus, any tuple in $F^t$ for which $\tau$ is a rejecting transcript must be in $G^t$. This implies that 
\[|G|^t\geq |F|^t\cdot 2^{-c} \implies |G|\geq |F|\cdot 2^{-c/t}.\]
Include the transcript $\tau$ in $A_n$ and repeat this step by taking $F=F\setminus G$ until there is no more $x\in \overline{L}$ that is not covered by some $\tau$ in $A_n$.

There are at most $2^n$ instance of $x$ of bitlength $n$. It is easy to see that by repeating the above procedure $\ell$ times, the set $A_n$ covers at least $(1-(1-2^{-c/t})^\ell)\cdot 2^{n} \geq (1- e^{-\ell \cdot 2^{-c/t}})\cdot 2^{n} $ inputs. It implies that all instances will be covered after $\mathcal{O}\left(\frac{n}{2^{-c/t}}\right)$ repetitions of the above step. As $c=\mathcal{O}(t\log t)$ where $t$ is some polynomial in $n$, we get that the size of the set $A_n$ is $\poly(n)$. Notice that each transcript is also polyonomially bounded in $n$ and the running time is also bounded in $\poly(n)\cdot T$. The resulting algorithm for $\overline{L}$ runs in $\NTIME(\poly(n).T)/\poly$.     
\end{proof}

\GenOracleCom*
\begin{proof}
Let's assume that there is a randomized oracle communication protocol of $k$-$\sat$ of cost $\mathcal{O}(n^c)$ for some constant $c<k$ and runs in time $T$. From Lemma~\ref{lem:VCover-SAT}, it gives  an oracle communication protocol for $k$-Vertex cover of cost $\mathcal{O}(n^c)$. Notice that, by a trivial reduction from $k$-Clique to $k$-Vertex cover, it gives a $\mathcal{O}(n^c)$ cost protocol for $k$-Clique. Therefore, by Lemma~\ref{lem:OR-SAT-Clique}, there is an oracle communication protocol for OR($3$-$\sat$) of cost $\mathcal{O}\left(\left(n\cdot \max\{n,t^{1/k+o(1)}\}\right)^c\right)$. As $c$ is a constant less than $k$, if we take $t$ as sufficiently large polynomial in $n$ it gives a $\mathcal{O}(t\log t)$ cost randomized oracle communication protocol for OR($3$-$\sat$). Hence, if $\coNP \not\subseteq \mathsf{NTIME}(\poly(n)\cdot T)/\Poly$, then we get a contradiction from Lemma~\ref{lem:compression-implication}. It completes the proof.
\end{proof}

\end{document}